# *In Situ* Geochronology for the Next Decade: Mission Designs for the Moon, Mars, and Vesta

Barbara A. Cohen, *NASA Goddard Space Flight Center, Greenbelt MD*
Kelsey E. Young, *NASA Goddard Space Flight Center, Greenbelt MD*
Nicolle E. B. Zellner, *Albion College, Albion, MI*
Kris Zacny, *Honeybee Robotics, Pasadena CA*
R. Aileen Yingst, *Planetary Science Institute, Tucson AZ*
Ryan N. Watkins, *Planetary Science Institute, Tucson AZ*
Richard Warwick, *Lockheed Martin Engineering, Littleton, CO*
Sarah N. Valencia, *University of Maryland / NASA Goddard Space Flight Center, Greenbelt, MD*
Timothy D. Swindle, *Lunar and Planetary Laboratory, University of Arizona, Tucson AZ*
Stuart J. Robbins, *Southwest Research Institute, Boulder, CO*
Noah E. Petro, *NASA Goddard Space Flight Center, Greenbelt MD*
Anthony Nicoletti, *NASA Goddard Space Flight Center, Greenbelt MD*
Daniel P. Moriarty, III, *University of Maryland / NASA Goddard Space Flight Center, Greenbelt, MD*
Richard Lynch, *NASA Goddard Space Flight Center, Greenbelt MD*
Stephen J. Indyk, *Honeybee Robotics, Pasadena CA*
Juliane Gross, *Rutgers University, Piscataway NJ*
Jennifer A. Grier, *Planetary Science Institute, Tucson AZ*
John A. Grant, *National Air and Space Museum, Smithsonian Institution, Washington, DC*
Amani Ginyard, *NASA Goddard Space Flight Center, Greenbelt MD*
Caleb I. Fassett, *NASA Marshall Space Flight Center, Huntsville, AL*
Kenneth A. Farley, *California Institute of Technology, Pasadena CA*
Benjamin J. Farcy, *University of Maryland, College Park, MD*
Bethany L. Ehlmann, *California Institute of Technology, Pasadena CA*
M. Darby Dyar, *Planetary Science Institute, Tucson AZ*
Gerard Daelemans, *NASA Goddard Space Flight Center, Greenbelt MD*
Natalie M. Curran, *Catholic University of America / NASA Goddard Space Flight Center, Greenbelt, MD*
Carolyn H. van der Bogert, *Institut für Planetologie, Westfälische Wilhelms-Universität, Münster, Germany*
Ricardo D. Arevalo, Jr, *University of Maryland, College Park, MD*
F. Scott Anderson, *Southwest Research Institute, Boulder, CO*

## ABSTRACT

Geochronology, or determination of absolute ages for geologic events, underpins many inquiries into the formation and evolution of planets and our Solar System. Bombardment chronology inferred from lunar samples has played a significant role in the development of models of early Solar System and extrasolar planetary dynamics, as well as the timing of volatile, organic, and siderophile element delivery. Absolute ages of ancient and recent magmatic products provide strong constraints on the dynamics of magma oceans and crustal formation, as well as the longevity and evolution of interior heat engines and distinct mantle/crustal source regions. Absolute dating also relates habitability markers to the timescale of evolution of life on Earth. However, the number of geochronologically-significant terrains across the inner Solar System far exceeds our ability to conduct sample return from all of them. In preparation for the upcoming Decadal Survey, our team formulated a set of medium-class (New Frontiers) mission concepts to three different locations (the Moon, Mars, and Vesta) where sites that record Solar System bombardment, magmatism, and/or habitability are uniquely preserved and accessible. We developed a notional payload to directly date planetary surfaces, consisting of two instruments capable of measuring radiometric ages *in situ*, an imaging spectrometer, optical cameras to provide site geologic context and sample characterization, a trace element analyzer to augment sample contextualization, and a sample acquisition and handling system. Landers carrying this payload to the Moon, Mars, and Vesta would likely fit into the New Frontiers cost cap in our study (~$1B). A mission of this type would provide crucial constraints on planetary history while also enabling a broad suite of investigations such as basic geologic characterization, geomorphologic analysis, ground truth for remote sensing analyses, analyses of major, minor, trace, and volatile elements, atmospheric and other long-lived monitoring, organic molecule analyses, and soil and geotechnical properties.

# 1. Introduction

Geochronology, or the determination of absolute ages for geologic events, underpins many inquiries into the formation and evolution of planetary bodies and our Solar System. Bombardment chronology inferred from lunar samples has played a significant role in the development of models of early Solar System and extrasolar planetary dynamics, as well as the timing of volatile, organic, and siderophile element delivery. Improvements and expansion of known absolute ages would yield new insights into many scientific questions. For example, absolute ages of ancient and recent magmatic products provide strong constraints on the dynamics of magma oceans and crustal formation, as well as the longevity and evolution of interior heat engines and distinct mantle/crustal source regions. Absolute dating also relates habitability markers to the timescale of evolution of life on Earth. In addition, terrestrial laboratory radiometric and cosmic ray exposure dating of lunar samples, paired with crater size-frequency distributions (CSFDs) of the geologic units associated with the samples, have enabled the calibration of lunar cratering chronology functions. These functions allow the determination of model ages for unsampled geological units across the Moon and have been adapted for application on other terrestrial planetary bodies for which no samples of known provenance currently exist. Major advances in planetary science can thus be driven by geochronology in the next decade, calibrating body-specific chronologies and creating a framework for understanding Solar System formation, the effects of impact bombardment on life, and the evolution of planetary surface environments and interiors.

Absolute ages for formative events in the timelines of multiple worlds were a desire in both the 2003 and 2013 Planetary Science Decadal Surveys. For example, *Vision and Voyages* (2013) advocates efforts to "Determine the chronology of basin-forming impacts and constrain the period of late heavy bombardment in the inner Solar System and thus address fundamental questions of inner solar system impact processes and chronology." However, given the limitations of then-existing technologies, only sample return was considered a viable method for geochronology and was therefore used as a driver for recommending and implementing sample return missions such as OSIRIS-REx and Mars Sample Return. *Visions and Voyages* (2013) also recommended a New Frontiers-class mission to return samples from the Moon's South Pole-Aitken Basin to address the bombardment history of the inner Solar System. Such a mission was proposed multiple times by several groups (Duke 2003; Jolliff et al. 2012; Jolliff et al. 2017), and selected twice for Phase A study, but no mission concepts have yet been selected for implementation. Though there are many factors in play when making mission selections, sample return within the New Frontiers program seems to be accomplishable from small bodies (e.g., OSIRIS-REx). There may be additional costs and technical risks in realizing a robotic sample return mission to bodies with significant gravitational wells or those requiring supplemental mission elements (such as a communications satellite to enable lunar farside communication). Further challenging the sample return model, the number of chronologically significant geologic terrains across the inner Solar System far exceeds our financial capacity and projected technological ability to collect, cache, and return samples from all of them, much less do the same across the outer Solar System.

The recommended mission lists in both prior Decadal Surveys reflect the reality that for those decades, sample return was regarded as the only way to provide reliable and interpretable geochronologic constraints on planetary bodies. However, as a community, we now routinely use *in situ* geochemical (e.g., APXS, ChemCam) and isotopic techniques (e.g., mass spectrometers). While these may yield less sensitive or precise measurements than terrestrial laboratories, they are still sufficient to resolve major science questions associated with planetary environments throughout the Solar System where sample return is prohibitive due to financial, technical, and/or other challenges. In the last two decades, NASA has also invested significantly in the development of innovative *in situ* dating techniques. Instrument maturation programs (i.e., Planetary Instrument Concepts for the Advancement of Solar System Observations (PICASSO), Maturation of Instruments for Solar System Exploration (MatISSE), and Development and Advancement of Lunar Instrumentation (DALI)) will have brought the technology readiness levels (TRL) of instruments that can access complementary radiogenic isotopic systems (K-Ar and Rb-Sr) to TRL 6 by the time of the next Decadal Survey. Sample collection and handling systems have also matured, while informed/autonomous operational scenarios continue to evolve.

In preparation for the 2023 Planetary Science Decadal Survey, NASA commissioned several mission concept studies, selected via the Planetary Mission Concept Studies (PMCS) program. As one of these studies, our study was designed to formulate dedicated *in situ* Geochronology mission concepts to address science objectives for the Moon, Mars, and small bodies such as Vesta. The aim of this study was to investigate the viability of *in situ* dating missions to accomplish longstanding geochronology goals within a New Frontiers cost envelope. To accomplish this, our study team identified science goals and objectives, formed a notional payload, examined potential landing sites, and developed a spacecraft architecture for each destination. Our full PMCS report, including extensive details on the payload, spacecraft bus, and cost and schedule, can be found in Cohen et al. (2020).

# 2. Geochronology Mission Science Goals and Objectives

Characterizing the timing and relationships between geological processes across the Solar System is a major goal of planetary science. Our existing understanding of inner Solar System chronology is rooted in defining geologic epochs on each body, and then assigning absolute ages to those epochs by scaling the lunar production and chronology functions to different planetary conditions (Strom & Neukum 1988; Hartmann & Neukum 2001; Neukum et al. 2001; Schmedemann et al. 2014; Hiesinger et al. 2016a). Unfortunately, the lunar cratering record itself is unconstrained prior to the apparent bombardment of 3.9 Gyr ago (basin-forming epoch) and suffers from a roughly billion-year uncertainty between 1 and 3 Ga ("middle ages"; Hartmann et al. 2000; Ryder et al. 2000; Stöffler et al. 2006; Chapman et al. 2007; Fassett & Minton 2013; Robbins 2014; Bottke & Norman 2017). Geologic epochs on different planetary bodies have been defined by events that have little apparent relation to each other (Fig. 1). Therefore, refining the lunar crater chronology curve and calibrating body-specific chronologies is critically important for comparing planetary histories, contextualizing Solar System dynamics, and developing an interplanetary perspective on the evolution of planetary surfaces, interiors, and habitable environments. We chose three overarching Science Goals to investigate in this study:

**Goal A:** Determine the chronology of basin-forming impacts to constrain the time period of heavy bombardment in the inner Solar System, and thus address fundamental questions related to inner Solar System impact processes and chronology. Ascertaining the early flux of impactors on all planetary bodies across the inner and outer Solar System is necessary to understand the Solar System's dynamical evolution and processes occurring on nascent planetary bodies.

**Goal B:** Reduce the uncertainty for inner Solar System chronology in the "middle ages" (1-3 Ga) to improve models for planetary evolution, including volcanism, volatile evolution, and habitability. Lunar cratering chronologies are not well-calibrated in the period between 1 and 3 Ga, because there are no returned samples of known provenance with these ages. This deficit propagates into uncertainties in the chronology systems for the Moon and other bodies, resulting in large uncertainties in the history and duration of volcanic activity.

**Goal C:** Establish the history of habitability across the Solar System. Absolute ages of potentially habitable terrains would help resolve when localized environments within the inner Solar System could have supported biological activity.

From these overarching objectives, we formulated targeted Science Objectives that would lead to progress on these goals by *in situ* investigations. To do this, we needed to adopt a quantitative measurement requirement. The 2015 NASA Technology Roadmap calls out *in situ* dating as an important investment, suggesting a minimum precision better than ±5% for rocks 4.5 billion years old (Ga) (approximately ±200 Myr, 2$\sigma$), and a desired precision of ±1% for 4.5 Ga rocks (or about ±50 Myr, 2$\sigma$). For this study, we used this precision to develop specific cases where this level of uncertainty would resolve Science Objectives specific to the Moon, Mars, and Vesta, tracing to LEAG, MEPAG, and SBAG goals documents (Table 1).

**Objective 1: Establish the chronology of basin-forming impacts by measuring the radiometric age of samples directly sourced from the impact melt sheet of a pre-Imbrian lunar basin.** The leading model for lunar impact history, which is still under debate (Bottke & Norman 2017; Hartmann 2019), includes a pronounced increase in large impact events around 3.9 Ga (Tera et al. 1974; Ryder 1990). The factors that led to this "terminal lunar cataclysm" would likely have also led to large impacts throughout the inner Solar System as a "late heavy bombardment," influencing the habitability potential of Earth and other bodies. Recent work on lunar samples has identified the possibility that lunar sample collections may be biased by repeated (albeit unintentional) sampling of Imbrium basin ejecta at the various Apollo landing sites (Schaeffer & Schaeffer 1977; Haskin et al. 1998; Cohen et al. 2000; Norman et al. 2010). Few rocks older than 3.9 Ga exist on the Earth, so samples from the basin-forming epoch are scarce, but this bombardment would have occurred on an early Earth with an atmosphere, oceans, and continents, and may have influenced the course of biologic evolution (Maher & Stevenson 1988; Mojzsis &

Harrison 2000). Large impacts may have had a similar influence on potential biologic evolution on other planets (e.g., Mars or even Venus), early in their history.

Dynamical models that support an early bombardment of the Moon assume lunar bombardment is strongly linked to the broader processes describing the endgame of planet formation (Morbidelli et al. 2018). A successful model must not only explain what is found on the Moon, but also constrain early bombardment on Mercury, Earth, Mars, the asteroids, and potentially bodies in the outer Solar System as well. These models, which extend from the gas-dust dynamics of forming disks to giant planet migration, are invoked to understand our Solar System as well as systems of exoplanets around other stars. As we seek to better link what we know about these other systems, we are left with a fundamental question: Is our Solar System typical or anomalous? One of the best ways to address this question is to determine what processes occurred and their timing and duration in the early Solar System, and then compare these findings to what is observed in planetary systems that are currently forming around other stars. A key test of these dynamical models is whether the terrestrial planets and asteroid belt experienced a relative "lull" in impacts between formation and later bombardment.

Geologic observations of surface morphologies and geophysical data of the Moon have revealed more than 50 distinct basins and possibly more candidate basins whose surface expressions have presumably been obscured by subsequent impact resurfacing (Wilhelms 1987; Spudis 1993; Frey 2011; Featherstone et al. 2013; Neumann et al. 2015). Cross-cutting relationships of ejecta and crater densities allow reconstruction of a relative time-sequence of the basins that have a clear surface expression. Recent work improving the extraction of CSFDs from heavily cratered terrains has allowed fitting of basin deposits with absolute model ages (Orgel et al., 2018). However, the precision of these fits is unknown due to a paucity of calibration points in this part of the lunar cratering chronology (e.g., van der Bogert and Hiesinger, 2020). Wilhelms 1987; Fassett et al. 2012). In large part, this reflects our inability to link individual lunar samples with specific basins or craters with a high degree of confidence, even when carefully collected by astronauts. Stöffler and Ryder (2001) provide a comprehensive summary of the radiometric dates of lunar samples available at that time and their interpretation of the ages of key nearside basins such as Imbrium, Serenitatis, Crisium, and Nectaris. However, more recent work has called into question the geologic interpretation of samples from Serenitatis, Crisium, and Nectaris, pulling the pin on the only constraints on a lunar cataclysm (Norman, 2009; Norman et al. 2010; Spudis et al. 2011; Zellner 2017).

There is general agreement in the community that Imbrium formed at 3.9 Ga, though the exact age depends on the geologic sample, radiometric dating system, and interpretation – ranging from 3.86±0.09 Ga to 3.91±0.09 Ga to 3.934 ± 0.03 Ga (Deutsch & Stöffler 1987; Stöffler & Ryder 2001; Norman et al. 2010; Merle et al. 2014). These differences are too small to resolve with *in situ* dating; instead, we seek a basin age that could be distinguished from 3.9 Ga to serve as a second pin in lunar basin stratigraphy. The key to a cataclysm is the duration of bombardment. Even in a case where results might be in between the two extremes (for example, 4.0 ± 0.2 Ga), we can use that interval to infer a moderately strong form of the cataclysm. In addition to measurement precision, careful site selection and sample characterization are keys to remote identification of impact-melt samples derived from a pre-Imbrian basin and distinguishing them from KREEP-rich Imbrium samples. Finally, once we identify the impact melt-derived material *in situ*, characterize its elemental composition and mineralogy (for geological context), and determine its age, we may be able to find the same materials in the returned sample collection and/or lunar meteorite inventory. Such samples could then be used to measure a more precise age on the same lithology in a terrestrial laboratory, leveraging *in situ* geochronology to acquire returned-sample-quality data.

*In situ* dating precision of ±200 Myr with 95% confidence (2$\sigma$) may be sufficient to date the impact-melt sheet of lunar basins thought to be significantly older than the Imbrium basin, which would place them either within the canonical cataclysm (3.9 Ga) or as part of a declining bombardment in which most impacts are 4.2 Ga or older.

**Objective 2: Establish the age of a very young lunar basalt to correlate crater size-frequency distributions with crystallization ages.** Calibration of the post-basin epoch lunar chronology is based on Apollo and Luna samples from lunar mare basalt flows and younger benchmark craters where samples yield radiometric formation and exposure ages, including Copernicus, Tycho, North Ray, and Cone craters (Stöffler & Ryder 2001). Higher-resolution imaging of the Moon made available by the Lunar Reconnaissance Orbiter, Kaguya, and Chang'E-1 has been used to update the crater size-frequency distribution (CSFD) ages of key lunar geological units for the calibration of the cratering chronology and to check and improve existing crater statistics (Hiesinger et al. 2003; Haruyama et al. 2009; Hiesinger et al. 2012; Robbins 2014; Williams et al. 2014b; Wang et al. 2015; Hiesinger et al. 2016b). Detailed CSFD measurements have also been made for the selection of young basalt units that could be favorable landing sites for sample return or *in situ* measurements (Hiesinger et al. 2003; Qian et al. 2018; Stadermann et al. 2018). However, different authors' chronologies yield a large range of possible absolute model ages for units with cumulative crater frequencies that place them in the middle part of the chronology function (van der Bogert & Hiesinger 2020). For example, crater density results differ by a factor of 2-3 for the area

surrounding North Ray crater. As a result, a revised chronology by Robbins (2014) would predict that a lunar surface previously dated at 3 Ga may have an updated model crater age as young as 1.9 Ga.

Under the "classic," sample-based lunar chronology, mare volcanism is thought to have reached its maximum volumetric output between 3.8 and 3.2 Ga (Shearer & Papike 1999). New crater-density observations imply peak volcanism may have extended for an additional billion years through 2.5 Ga (Hiesinger et al. 2000; Braden et al. 2014; Qiao et al. 2017). Such a finding would dramatically revise our understanding of the thermal evolution of the lunar mantle, the abundance and distribution of radioactive heat-producing elements (e.g., K, U, Th), and release of indigenous lunar volatiles (Needham & Kring 2017). One or more *in situ* age measurements of lunar basalts significantly younger than 3.2 Ga would greatly improve the calibration of lunar cratering chronologies within this time frame. Not only would this refine our understanding of the duration of lunar mare basalt volcanism, but the improved lunar calibration would propagate forward to improved chronologies for other planetary bodies.

*In situ* dating precision of ±200 Myr with 95% confidence ($2\sigma$) would be sufficient to reduce the uncertainty in absolute model ages derived from crater size-frequency distribution measurements to no more than 20% of the current uncertainty shown between different lunar chronology functions.

**Objective 3: Establish the age of a well-exposed Hesperian martian igneous terrain to correlate crater size-frequency distributions with crystallization ages.** Ages for ancient and recent magmatic products provide strong constraints on the dynamics of magma oceans and crustal formation, the longevity and evolution of interior heat engines and mantle/crustal source regions. On Mars, the "middle ages" are geologically rich, including the cessation of abundant volcanism and the formation of hydrated minerals (Ehlmann et al. 2016a). Revision of the middle Mars chronology might imply that these processes could have lasted for a billion additional years, revising models for martian thermal evolution and allowing a longer era of abundant volatiles and potential habitability, or might demonstrate that the existence of habitable conditions was relatively brief. Given the current uncertainties in the absolute age of volcanic surfaces of middle and especially younger martian age, there is a large uncertainty in the history and duration of interior processes on Mars. While InSight is shedding light on the current state of the martian interior, the path to the present remains poorly constrained. For example, a factor of two uncertainty in the absolute ages of Hesperian lava plains has large implications for the source of internal heat responsible for driving interior processes through time. Increasing uncertainty in the absolute ages for the Amazonian results in even less awareness of how interior processes and associated igneous activity evolved over the past ~2-3 Ga. Critical advances in understanding planetary volcanism at benchmark igneous provinces, coupled with elemental and mineralogical analyses, would provide geologic context and critically distinguish mantle sources.

The Mars Sample Return program represents an extensive effort to bring back samples from materials within Jezero Crater. The selected *Perseverance* landing site, Jezero crater, contains an areally extensive, well-cratered unit overlying basin fill material and embaying delta outcrops, though the origin of the unit as volcanic or sedimentary is not yet established (Goudge et al. 2018; Rogers et al. 2018; Shahrzad et al. 2019). Investigations with the rover payload will determine if this unit is volcanic, volcaniclastic, or sedimentary; returned samples from the unit may determine a crystallization age that would be relatable to the martian crater flux function, or the formation and exposure age of its detrital precursors. Evaluating these possibilities will begin in early 2021 as *Perseverance* lands and begins investigating. However, even if the Jezero samples represent a major step forward in correlating absolute ages and relative crater density, there is a bigger, broader issue in knowing the systematic factors that relate martian crater spatial density to age throughout its history. By analogy to the lunar curve, multiple data points on Mars are needed to provide a reference frame for understanding whether scaling factors between the Moon and Mars are correct and how the changing Solar System dynamical environment affected both planets.

*In situ* dating precision of ±200 Myr ($2\sigma$) is sufficient to radically improve our understanding of Mars' volcanic history for geodynamics and interior cooling, assign widely-separated igneous provinces absolute ages, and examine the compositional progression of igneous sources with time. Perhaps most importantly, such dating precision should pin the absolute ages associated with geologic history in middle and late martian history, thereby significantly reducing the current ~2× uncertainty in the timing of late aqueous activity and the persistence of past habitability conditions. This would also provide a direct test of the quality of the adaptation of the lunar cratering chronology to Mars.

**Objective 4: Establish the epoch of martian habitability by measuring the radiometric age of Noachian clay-bearing stratigraphies.** Incomplete knowledge of absolute martian geochronology limits our ability to understand the timing of martian evolutionary milestones (Doran et al. 2004; Tanaka et al. 2014), specifically when Mars changed from a habitable environment to its present state (Ehlmann et al. 2016a). For example, the absolute age of the Noachian-Hesperian boundary is unknown. When did a warm, wet Mars become arid (Bibring et al. 2006), and did it occur before, after, or concurrent with the Imbrium impact on the Moon and the oldest intact rocks on Earth? Thus, martian climate change cannot yet be put into the context of Solar System history and the evolution

of life on Earth. Results from the *MSL Curiosity* rover mission have demonstrated that aqueous habitable settings existed on Mars in surface environments well into the Hesperian and maybe later (Grotzinger et al. 2015; Martin et al. 2017; Cohen et al. 2019). Yet, most models for the evolution of the martian climate suggest conditions were likely not favorable for surface habitable environments in much of the Hesperian, much less into the Amazonian. Knowing whether such environments could be older or younger than predicted in the current chronology is critical to understanding how the habitability of the planet evolved over time. Absolute age constraints with the precision of *in situ* dating would help resolve the large uncertainty in when Mars changed from a habitable environment to its present state. Importantly, they would allow the extension of this chronology to other parts of the planet via crater density statistics.

*In situ* dating precision of ±200 Myr (95% confidence, or 2$\sigma$) is sufficient to constrain the timing of Noachian unit formation. This would provide an important anchor for crater spatial densities of terrains hosting geologic evidence from Mars' most habitable period. The broadest epoch boundaries for the Late Noachian period span ~3.9-3.6 Ga (Michael, 2013) in different crater chronology models. For example, resolving the age of a late Noachian clay-bearing unit would test whether crater spatial densities from Mars have been correctly modeled. This has concomitant implications for reconciling the timing of the development of life on Earth (~3.5 Ga age for the oldest confirmed fossil evidence) and hypothesized spikes in early impact bombardment (~3.9 Ga).

**Objective 5: Establish the radiometric ages of vestan samples with well-established provenance.** The Main Asteroid Belt represents a large reservoir of bodies that have largely been devoid of endogenic processing since very early in Solar System history. However, they have experienced intense, often disruptive impact events, some of which contributed to the flux of material that impacts the inner planets. The asteroids also serve as probes of the dynamical history of the Solar System, via impact ages of meteorites derived from asteroid parent bodies (Bogard 2011; Jourdan 2012; Swindle et al. 2014). H-chondrites show a prominent group of ages between ~3.5 and 4.0 Ga, and eucrites and howardites derived from Vesta cluster at ~3.5 and 3.8–4.0 Ga (Bogard 2011; Cohen 2013; Kennedy et al. 2019). Connecting these ages to chronology of the Main Belt relies either on dynamical models or, for the few asteroids that have been visited by spacecraft, on model crater retention ages based on extrapolations of the lunar chronology curve applied to a set of bodies in a different dynamical environment.

One of the best targets for calibrating timescales is Vesta, the second-most massive object in the Main Belt. Vesta has a basaltic crust, and basaltic meteorites linked to Vesta by the Dawn mission and by earlier spectroscopic work have crystallization ages within a few million years of the birth of the Solar System. Hence Vesta is an object that formed early and then survived intact until the present, serving as a witness to all that has occurred in the Main Belt since more than 4.5 Ga. The major basin ages on Vesta have wildly different inferred ages, depending on which lunar chronology curve is applied. The giant basins Rheasilvia and Veneneia may be nearly as old as the large lunar basins, with the formation of Rheasilvia proposed at 3.4±0.1 Ga and Veneneia somewhat older (Schmedemann et al. 2014), or possibly as young as 1-2 Ga (Schenk et al. 2012). Meanwhile, most radiometric ages for meteorites derived from Vesta are >3.47 Ga (Bogard 2011; Cohen 2013; Kennedy et al. 2013), although there are some feldspar grains with ages only slightly older than 1 Ga (Lindsay et al. 2015; Jourdan et al. 2017).

*In situ* dating precision of ±200 Myr (95% confidence, or 2$\sigma$) is more than sufficient to constrain Vesta's geologic timescale by dating key stratigraphic craters and contiguous geologic terrains. Given the large disparity in ages derived by different logical assumptions, this level of precision would not only reveal the ages of key basins but would also set firm constraints on the impactor flux estimates used throughout the Main Asteroid Belt. These measurements would also provide a direct test of the quality of the adaptation of the lunar versus asteroid cratering chronologies to Vesta and the rest of the Main Asteroid Belt.

# 3. Geochronology Mission Science Requirements and Payload

A Science Traceability Matrix (STM; Table 1) codifies the measurement and mission requirements needed to address the geochronology Science Goals and Objectives and connects measurements to the lander payload. The overarching measurement requirement in all cases is to understand the age of the desired lithology. To accomplish this, several measurements and associated observations need to be made, including a) using radiometric chronology to directly measure the age of samples derived from the target lithology with precision better than or equal to 5% of the lithology age (95% confidence, or 2$\sigma$); b) contextualizing the desired lithology using petrology, mineralogy, and/or elemental chemistry; and c) relating the measured lithology age to crater counting of the lithology's terrain.

For this study, measurement requirements for all goals and objectives would be met by carrying a single notional payload comprising representative instruments, all of which have substantial development and heritage for this type of mission (Table 2). A notional payload would conduct *in situ* geochronology on samples derived from the surface under investigation and provide context to the samples and ages by mapping mineralogy (e.g., olivine, pyroxene, iron oxides, plagioclase, and aqueous alteration minerals including phyllosilicate, sulfate, carbonate, and other hydrated salts). It would image the samples at the microscale to determine grain size distributions, textural relationships, etc. and measure the major- and trace-element geochemistry of the samples to establish parentage and trace geological processes. Finally, it would constrain the lithologic evolution and contribute to understanding the geology of the landed site and its lithologic units, relating it to maps and crater counts determined from remote sensing.

To make significant advances in creating a geochronologic framework, *in situ* geochronology must yield ages that are both precise and accurate – that is, the measurement techniques must yield small uncertainties on the calculated age, and that age must be recognizable and interpretable as a geologic event. Radiometric dating, or the process of determining the age of rocks from the decay of their radioactive elements, has been in widespread use for over half a century. Dating techniques require measuring the parent and daughter isotopes in a pair (for this study, $^{86}$Rubidium-$^{86}$Strontium, or Rb-Sr, and $^{40}$Potassium-$^{40}$Argon, or K-Ar) to determine when a rock closed to addition or loss of its radioactive elements or their decay products. Many rocks are amenable to Rb-Sr and K-Ar dating in terrestrial labs, including igneous rocks, phyllosilicates/clays, and sulfates. Each mineral can record a different event in the rock's history, from initial crystallization to alteration events such as impact or weathering.

Multiple groups have made substantial progress on bringing radiometric dating techniques to flight implementation; a comprehensive review of developments and proposals appears in Cohen et al. (2019). Armed with this knowledge, we are able here to narrow our consideration to a subset of potential instruments, their precision, and potential for implementation. For many planetary materials, it may be possible to measure ages using more than one system, which is common practice in terrestrial laboratory geochronology. Agreement between multiple chronometers increases confidence in the interpretation of the geologic events experienced by the sample, though disagreement does not negate the inherent value of each measurement. We therefore baselined two independently-developed *in situ* dating instruments that together can access both the Rb-Sr and K-Ar radiometric systems.

The Chemistry and Dating EXperiment (CDEX; Fig. 2a) is a Laser Ablation - Resonance Ionization Mass Spectrometer (LA-RIMS) developed via the NASA PIDDP and MatISSE Programs (Anderson et al. 2015a; Anderson et al. 2015b). CDEX uses LA-MS to obtain elemental abundances and LA-RIMS to obtain isobar-free Rb-Sr dates, in addition to Pb-Pb dating currently in development. Because CDEX uses an isochron approach, scanning the laser beam over the sample to map elemental abundances at microscopic scales, it can provide context with which to interpret isotopic age data, for example by recognizing secondary alteration. CDEX is based on two prototypes and designs proposed for Mars *Perseverance* and Discovery 2015. The first generation CDEX system has been used to demonstrate Rb-Sr and Pb-Pb measurements on terrestrial analogues and Martian meteorites with precision better than 200 Myr (Anderson et al. 2015a; Anderson et al., 2015. 2015b; Anderson et al. 2020). CDEX uses laser ablation to vaporize a small sample of the target rock, generating >99.9% neutral atoms. Sr is selectively ionized by using lasers tuned to electronic resonances (461 and 554 nm for Sr) in the neutral atoms, followed by photoionization of the excited atoms with a 1064-nm laser. This process for Sr is followed a couple of microseconds thereafter by the corresponding process for Rb (using 780 and 776 nm for the resonances) in the same ablation plume. This staggered ionization offsets the arrival of Sr from Rb ions at the detector of a time-of-flight mass spectrometer, eliminating isobaric interferences between Rb and Sr, and ensuring that the atoms come from the same ablation event. For Pb resonance ionization, the ablated plume is illuminated with lasers tuned to the 283.3-nm and 600.2-nm resonances and use the same 1064-nm light for photoionization. CDEX typically measures 100-300 locations on a sample in a raster pattern, ablating 50-μm diameter laser spots, thus sampling a range of different minerals for Rb-Sr isotope ratios. After every fifth spot, CDEX measures a well-characterized standard in the same manner. Spots with an isotope signal-to-noise ratio (SNR) > 2 are kept for the final analysis. When mapped onto a photomicrograph, the magnitude of the $^{87}$Rb/$^{86}$Sr ratio derived from corrected spectra highlights the different minerals in the sample, commonly matching with elemental abundance maps produced in LA-MS mode. The CDEX RI lasers can be turned off to map elemental chemistry in LA-MS mode, or measure organics in two-step laser mass spectrometry (L2-MS) mode.

Several laboratories have developed breadboards that provide comprehensive compositional analysis, including mineralogical identification, major element chemistry, and K-Ar dating, of solid samples by measuring the parent K via laser induced breakdown spectroscopy (LIBS), and daughter Ar via mass spectrometry (Cohen et al. 2014; Devismes et al. 2016; Cho & Cohen 2018; Cattani et al. 2019). Of these prototypes, the KArLE investigation developed through the NASA PIDDP and DALI programs is the most mature (Fig. 2b); in fact, the core technologies of this instrument have already been demonstrated on the *Curiosity* mission (Farley et al. 2014; Sautter et al. 2014; Conrad et al. 2016; Le Deit et al. 2016; Cohen et al. 2019) and proposed for Mars *Perseverance*. KArLE completely releases Ar from small pits using laser ablation and admits the released gas to the mass spectrometer. Using laser ablation enables the technique to be applied to solid, unprepared samples such as chips or pebbles rather than crushed/processed powders, and surface contamination can be ablated away without interfering with the chemical analysis. KArLE also permits multiple laser measurements to be made on a single sample, creating a linear array of measurements with a slope proportional to the age of the rock (an internal isochron). Using multiple measurements to construct an isochron decreases the uncertainty in the inferred age and increases the robustness of the interpretation. The isochron approach also obviates the need to independently assume or determine any initial or trapped contributions to $^{40}Ar$ in a bulk sample. The LIBS-MS family of instruments is promising for near-term implementation because its components (LIBS, MS, and cameras to measure the ablation pit volume) have successfully flown aboard the *Curiosity* and *Rosetta* missions. Quantification of elements by LIBS and the volume measurement by optical metrology are relatively imprecise compared with mass spectrometry, leading to an estimated uncertainty using this technique of ±8-16% (2σ) in individual measurements. Reduced uncertainty in the inferred age may potentially be achieved using multiple-point isochrons, approaching the guidelines set out in the NASA Technology roadmap. Each element (LIBS, MS, camera) makes measurements beyond geochronology, providing microimaging, elemental analysis (including volatile elements like H and Cl), and volatile compound mass spectrometry. The same measurements also yield cosmic-ray exposure ages.

Geochronology analyses should be paired with other observations that provide context and further enhance the science return of a prospective mission. The uncertainty in a geochronology measurement could be influenced not only by technological capabilities but also by the complexity (e.g., mineralogy, alteration history, etc.) of the planetary material and geologic setting being investigated. Thus, sample selection, location, and geological context are just as important as the analytical methodologies that enable radiometric dating.

Both geochronology instruments detect a range of major and minor elements that would assist in associating samples with specific lithologies, but complementary trace element analysis (ppmw levels and below) would enable discrimination of genetic relationships between different planetary materials. In terrestrial laboratories, Inductively Coupled Plasma-Mass Spectrometry (ICP-MS) is the benchmark for the quantitative measurement of trace element abundances in solid and/or liquid samples. For example, basalt flows with common mineralogy and/or major element characteristics derived from distinct source regions may be distinguished by rare earth element (REE) abundance patterns, redox-sensitive trace element ratios (e.g., V/Sc or Ce/Ce*), or temperature-sensitive partition coefficients (e.g., Ni in olivine). Aqueous alteration can be tracked via the dynamics of fluid-mobile elements (e.g., solubility B > Ca > Ba), and exogenous materials identified by enrichments in siderophile elements (e.g., Mo/Ce) in impact melts. Further, enhanced science return from trace elements could include determining volatile element depletion (e.g., Na, Zn, and Pb) and/or refractory element enrichment (e.g., Al, Ti, and REE); evaluating distributions of heat-producing elements (e.g., K, Th, and U) in silicate minerals; and constraining fluid interactions and chemical weathering via Li isotopes. Notably, such instrumentation could provide access to additional geochronometers, such as the U-Th-Pb system, providing corroborative age measurements.

The adaptation of ICP-MS technology for spaceflight applications is currently being pursued through the NASA PICASSO Program. Low-pressure plasma pioneered through the NASA SBIR/STTR Program (Taghioskoui & Zaghloul 2016) is being integrated with a quadrupole mass spectrometer originally manufactured as an engineering unit of the Sample Analysis at Mars (SAM) investigation onboard the *Curiosity* rover (Mahaffy et al. 2012). A custom stack of ion optics, including a quad deflector in the image of the mass spectrometers flown on the *LADEE* and *MAVEN* missions (Mahaffy et al. 2015a; Mahaffy et al. 2015b), serves as an interface between the plasma and mass analyzer; it separates charged and neutral particles, maximizes ion transmission, and relaxes pumping requirements. Under these current investments, an ICP-MS instrument for planetary surfaces will achieve TRL 4 by 2021 through system-level breadboard demonstration and performance validation in a laboratory environment. Further maturation to TRL 6 would be required by mission PDR. This could be accomplished by a follow-on MatISSE or DALI investment to mature the ICP-MS independent of a flight mission, or a mission investment to mature the instrument. If the ICP-MS were to fail to achieve TRL 6 prior to mission PDR, it could be descoped at that time with a moderate loss to science, bearing in mind that major-and minor-element analysis would still be conducted by the other instruments in the payload. The ICP-MS would use a laser ablation system and mass spectrometer similar to those baselined for KArLE; for this mission concept study, the team decided to efficiently package the two instruments to conserve mass/volume and eliminate duplication of these components.

Geologic context would provide the framework necessary for interpreting the results of *in situ* geochronology. Verifying that selected samples were associated with cohesive surface units, rather than deposits not definitively representative of the specific locality under investigation, would further enable interpretation of radiometric ages and deduction of relationships to mapped surface features. Visible/color imaging of the ingested samples at hand-lens scales would yield information regarding lithology, grain characteristics (modal proportions, crystal habits, etc.), and petrology (e.g., textural relationships, such as overgrowths). Millimeter- to centimeter-scale images of the sampling location would provide an understanding of the surface components from which the sample would be acquired. Finally, panoramic imaging of the landing site would reveal the local outcrop features, textures and morphologies at mm to m-scale, yielding information about lithology and relationships, and facilitating reconstruction of the local and regional geology of a site. Color imaging, in particular, would capture differences in regolith materials that are expressed through subtle color variations (e.g., orange and black glasses visible to the Apollo 17 astronauts against the background of surrounding non-pyroclastic regolith). For this effort, the team included a suite of imagers that are part of Malin Space Science Systems' commercial ECAM product line (Maki et al. 2003; Maki et al. 2012; Maki et al. 2018), though suitable imagers have flown on a variety of planetary missions in diverse environments, including Mars (MAHLI, WATSON, ECAMs, Mastcam, and Pancam) and asteroids (ECAMs on OSIRIS-REx). The Heimdall investigation will use four of these imagers on a CLPS mission to the Moon in 2022 (Fig. 3a; Yingst et al. 2020). Therefore, all considered imagers will be TRL 9 and available for any mission in the next decade.

Infrared (IR) spectroscopy detects electronic and vibrational absorptions related to mineralogy in reflected light. The most important role of the IR spectrometer would be to identify samples of targeted lithologies, differentiating them from other materials. An IR imaging spectrometer would be particularly well-suited to this task to provide mineralogical composition with spatial context for every pixel in an image. Thus, at landscape scales, the IR spectrometer would identify and map at centimeter-scale discrete lithologic units to relate them to mapped lithologic units and crater densities of lithological units determined from orbital remote sensing. In the triage station, the IR spectrometer would resolve candidate pebble/cobble samples at millimeter-scale, allowing discrimination of sample domains by petrologic type. Infrared spectral data would be used to determine the diversity of lithologies present at the landing site and differentiate target lithologies for further measurement. Shortwave visible infrared imaging spectroscopy (SWIR; 0.6-3.6 μm) is most appropriate for this mission concept study because Mars dust obscures the rock in visible and near-infrared (VNIR) wavelengths, providing less compositional information (Murchie et al., 2019), and VNIR signatures on the Moon are modified by the effects of nanophase $Fe_0$ particles that are produced by space weathering (Pieters & Noble 2016). SWIR data from the IR imaging spectrometer would be used to identify different lithologies by mapping electronic and vibrational absorptions in reflected light that are related to mineralogy: olivine, pyroxene, ferric iron oxides, Fe plagioclase, and aqueous alteration minerals including phyllosilicates, opaline silica, sulfates, carbonates, and other hydrated salts. Thermal infrared data would be extremely powerful for this purpose as well (Christensen et al. 2001); however, orbital data available have insufficient spatial and spectral resolution to determine individual lithologies on Mars, the Moon, and Vesta. Several high TRL, low mass, high capability systems for SWIR imaging and point spectroscopy have been developed (Pilorget & Bibring 2013; Van Gorp et al. 2014; Cook et al. 2016; Ehlmann et al. 2019). The team chose for this study JPL's Ultra Compact Imaging Spectrometer (UCIS), an Offner spectrometer using e-beam gratings and HgCdTe detectors drawing direct heritage from the *Moon Mineralogy Mapper* $M^3$ instrument (Fig. 3b). UCIS collects reflectance spectra from 600-3600 nm, enabled by cryogenic cooling via a small Peltier cooler. UCIS was matured to TRL 6 for Mars and Vesta under MatISSE (Van Gorp et al. 2014) is now being adapted to TRL 6 for the Moon's challenging thermal environment through the DALI Program (Fraeman et al. 2020).

At present, geochronology measurements are not standoff or remote techniques; all share a common need for sample acquisition, manipulation, and analysis in a sealed and evacuated chamber to prevent neutral particles and ions liberated from the sample from escaping, and to enable quantitative measurements of parent and daughter isotopes and abundances. Therefore, a sample acquisition and sample handling system would be a required payload element. For this study, tens of individual samples would need to be acquired and dated for each science objective to achieve statistical confidence in the results. Sample manipulation of rocks using an arm or carousel is at TRL 9, having been used successfully on multiple missions including SAM on *Curiosity* (Mahaffy et al. 2012), and may be expected to be reasonably adaptable to other bodies and missions with only minor engineering modification. There are several methods that could acquire suitable rocks from various planetary surfaces, depending on the characteristics of the surface. For Mars, drilling capabilities have been extensively developed for the Mars *Perseverance* rover (as well as the *ExoMars* rover in 2022); core collection and insertion into a chamber has been laboratory tested (Zacny et al. 2014). For surface and near subsurface regolith, scoops and sieves with robotic arms have been used on missions such as *Apollo* and *Phoenix* (Arvidson et al. 2009). A robotic arm to collect and sieve regolith could provide access to a wider workspace around the lander, potentially mitigating circumstances such as landing in an area with very fine regolith or small-scale heterogeneity.

Alternatively, pneumatic-type systems are being developed that can loft and sort regolith into rocks and fines. Pneumatic sample delivery is currently at TRL 5-6 maturity, having been tested on an actual lander (Masten) and in vacuum chambers; such a system is also slated to launch on the JAXA *Mars Moons eXplorer* (MMX) mission in 2024 and on a CLPS lunar lander in 2022/2023, so it will have been fully developed for flight for these destinations. For this mission study, we selected Honeybee Robotics' PlanetVac system (Zacny et al. 2014) as a baseline, though any of the others could be considered with appropriate modifications to the concept of operations. PlanetVac (Fig. 4) comprises sampling cones, pneumatic nozzles inside the cone, compressed gas tanks, and pneumatic sample transfer lines. Pneumatics released from the nozzle would be directed at the surface, exciting the regolith and directing material through pneumatic plumbing. Samples would be transferred by the pressure differential caused by the released compressed gas and the environmental vacuum at the transfer lines' exhaust. Sieves in line with the cone and transfer lines would restrict the size of particles allowed to enter the transfer lines, preventing clogging and only allowing through rocks within the desired sampling size range. The samples of desired size would exit the PlanetVac system into a triage station, where they would be examined by onboard instruments, prioritized, and moved to individual instruments for analysis. The PlanetVac scheduled to launch to Mare Crisium in 2022/2023 has been designed to deliver regolith particles up to 1 cm in size. Vacuum chamber tests to date have shown successful delivery of these particle sizes.

# 4. Supporting Science

The uncertainty in a geochronology measurement is influenced not only by technological capabilities but also by the complexity (e.g., mineralogy, alteration history, etc.) of the planetary material and geologic setting being investigated. Thus, sample selection, location, and geological context are just as important as the analytical methodologies that enable radiometric dating. Verifying that selected samples are associated with cohesive surface units, rather than deposits not definitively representative of the specific locality under investigation, further enables interpretation of radiometric ages and deduction of relationships to mapped surface features.

## 4.1 Sampling statistics

The number of samples that need to be collected and analyzed is a function of a) how many aliquots of the lithology of interest are needed to ensure statistical confidence in assigning an age to the lithology; b) how much of the lithology of interest makes up the regolith at the landing site; and c) how many appropriately-sized rocks exist in the regolith.

The number of aliquots of any given lithology required for confidence in its measured age depends on multiple factors, including the accuracy and precision of the instrumental measurement and the cooperation of the samples themselves. Geochronology work in terrestrial laboratories takes these factors into account by running duplicates or triplicates of key samples to ensure that the sample behavior and age agree. The resultant quoted precision on the age is a combination of the measurement uncertainties and deviation among samples. For the science questions addressed by *in situ* geochronology missions to the Moon, Mars, and Vesta, measuring multiple aliquots of key samples increases the likelihood of obtaining robust age constraints and may overcome complications from overprinting impact events and other factors. Increasing the number of samples decreases the mean residual error and calculated standard deviation, potentially enabling the age of the geologic event to be interpreted with better precision than any individual measurement (e.g., terrestrial detrital mineral studies such as Coutts et al. 2019). Analyzing multiple aliquots would also ensure mitigation of secondary events that might be recognized in geochronology data, such as thermal or aqueous alteration. If results disagree among three aliquots of a lithology, the analysis shows that they were not all reset by the same event, or that there is some other complicating factor involved. The exact number of samples required to obtain robust statistical results depends on the exact question being asked. For this study, we adopted 10 samples as the requirement to achieve robust counting statistics that could improve the precision on the interpreted age by a factor of three if all samples yielded the same measured age.

Geologic setting and context are both crucial for missions intending to sample specific lithologies. Regolith formation on all bodies works vertically and laterally to mix materials to varying degrees, but geochemical signatures of different terrains are preserved, as observed from orbit. Sample return missions have shown that the most prominent component of any regolith sample is the substrate on which it formed, so samples collected from it would contain the lithology of interest. For example, the *Apollo 17* mission landed on the edge of the Serenitatis basin with the intent to sample Serenitatis ejecta and young basalt (Spudis & Pieters 1991). The Lunar Excursion Module landed directly on the basaltic surface of Mare Serenitatis, so there is little ambiguity about the origin of the high-Ti basaltic material seen in the returned samples (Fig. 5) – that is, landing on the basalt yielded samples of the underlying basalt (Rhodes et al. 1976). Though the regolith is developed primarily from the substrate, vertical and lateral mixing does occur. The disagreement in the community over whether the poikilitic impact melt breccias brought back from the ejecta-emplaced South Massif represent the Serenitatis impact (Dalrymple & Ryder 1996; Schmitt et al. 2017; Zhang et al. 2019) is an example of the complexities involved when interpreting the origin of samples from ejecta deposits. However, away from large basin ejecta sites, regolith mixing on the Moon and Mars is primarily local (not considering sedimentary processes).

In the Geochronology mission payload, we include instruments that would assist in distinguishing materials and associating them with remotely sensed data for the landing site and nearby units to enable identification of outliers that might be associated with fragments derived from more distant impact events. To further support science objectives, we have chosen candidate landing sites that are well-characterized from orbit and are geologically homogeneous with a lithology of interest that is clearly identifiable via remote sensing. For this study, we chose sites where our current understanding of remote sensing and geologic setting make it probable that the majority of the samples retrieved would represent the lithology of interest (see next section). Nonetheless, this assumption would need to be thoroughly vetted and refined for any proposed mission at any site. We conservatively include a factor of three on the sample estimates to account for potential mixing. This factor drives the number of samples that need to be collected and examined in triage.

Both geochronology instruments require rocks of 0.5-2 cm in diameter to conduct their analyses. To bound the total amount of sample that needs to be collected to yield enough rocks in this size fraction, we considered data from landing sites on the Moon and Mars and from a HED (vestan) meteorite (Table 3). It is worth noting here that robotic sampling of these smaller rocks, either vertically or laterally, is an excellent way to ensure sampling the complete lithologic diversity at any given site. Comparing small rocks separated from rake samples to samples carefully chosen by astronauts shows the same range of composition and frequency at the *Apollo 17* site (Fig. 5).

There are significant challenges to accurately estimating the frequency of 0.5-2 cm-sized rocks from returned lunar soil samples. Generally, lunar soil is defined as the size fraction <1 cm; therefore, samples were treated differently across the 1-cm boundary (McKay et al. 1991). The *Apollo 11* soils clumped together during sieving, so the recorded grain-size frequency distributions for the *Apollo 11* samples skew towards coarser fractions and are not representative of the lunar surface (Carrier 1973). For later missions, soils were passed through a 1-mm sieve before distribution to investigators for grain-size frequency distribution analyses, meaning that the recorded >1-mm size fraction data was variable. Aliquots of soil dedicated to grain size-frequency distribution analysis typically only included fragments <1 mm and reported size fractions as percentages, so the results cannot be extrapolated to larger grain sizes (Graf 1993).

Several researchers used photographs of the lunar surface to obtain the size-frequency distribution of boulder-sized rocks on the lunar surface. This type of image analysis is limited by image resolution (i.e., only boulders equal to or greater than the size of the pixels would be apparent). At the limits of the resolution of the image, where boulder size approaches pixel size, there is a roll-off in the number of countable blocks. Block counts at the *Surveyor* landing sites from orbital imagery follow a power law and rapidly roll-off as the resolution limit is approached (Fig. 6), which was ~2.5 m from *Lunar Orbiter* photography (Cintala & McBride 1995) and 0.5 m in *Lunar Reconnaissance Orbiter* images (Watkins et al. 2019), both of which are too coarse to confidently extrapolate to cm-sized rocks. A much smaller area (essentially within the radius of the lander footpads) allowed grain sizes to be counted down to 1 mm at the *Surveyor 1* and *3* sites using television images. These images encompass the grain size of interest and show thousands of fragments of appropriate size at the surface (Fig. 6; Shoemaker & Morris 1970). Their depth distribution is unknown, but if the surface density were extrapolated as uniform at depth, a volume of 0.6 L would be sufficient to yield 30 rocks 0.5 cm or larger (Table 3).

Because lunar soil statistics do not scale up and boulder counting does not easily scale down, we undertook a more accurate estimation of rock abundance with depth using data from the *Apollo* core samples. All *Apollo* missions used drive tubes to collect regolith core samples, representing a secular sampling of the upper ~0.5 m of regolith across different lunar geologic settings, which is analogous to material a robotic mission would be expected to sample. We identified individual fragments using core dissection diagrams of *Apollo* drive tubes (Fryxell & Heiken 1971; Lindsay et al. 1971; Allton 1978; Nagle 1979, 1980a, b; Ryder & Norman 1980; Fruland et al. 1982; Nagle 1982; Meyer 2016), counting all fragments with long axis dimensions 0.5-2 cm as a proxy for rock diameter (Table 4). The mission drive tubes used to collect the cores were 2 cm in diameter for *Apollo 11*, *12*, and *14* and up to 4 cm in diameter for the later missions. Historical core sketches only capture the axial plane of the core sample. Considering this geometry and the tube diameter, our method should capture most 0.5-2 cm rocks in the 2-cm cores but may undercount rocks in the larger, 4-cm diameter tubes. The 76001/2 double-drive tube currently being dissected under the ANSGA program will have a complete 3D CT scan that will make this method much more accurate. Among the 21 drive tubes and double drive tubes examined, the total number of rocks in the desired size range ranged from one to 72 (Table 4). Very mature soils have fewer large fragments while immature mare surfaces and landslide deposits contain more abundant, larger rocks. Surface age likely dominates this variability, because young surfaces have had less exposure to bombardment and are thus less broken down (i.e., grain sizes are larger). The number of fragments per volume using this method yields slightly lower abundances than simply extrapolating the surface count to a uniform distribution with depth. We adopted 0.6-1.2L as the amount of regolith that would yield 30 samples, including 100% margin on our estimates (Table 3).

The martian regolith is much different from the lunar case. The Moon has been subjected to billions of years of meteorite bombardment, creating a meters-thick, fine-grained regolith with little exposed bedrock. In contrast, the martian surface has been dominated by aeolian scouring and transport during the Amazonian, constantly exposing bedrock in some places and covering other areas with sediment. Previous mission landing sites have shown that in areas of active or recent deflation (e.g., the plains of Gusev crater, the lower flanks of Mt. Sharp in Gale crater), surfaces are covered by a scattered to continuous lag of coarse rock fragments too large to be transported by the wind. In many instances, fragment size distributions are as expected by impact fragmentation, with fragment composition consistent with derivation from local bedrock/materials (Grant et al. 2006). Because the plains often display a lag of fragments too large for aeolian transport, access to fragments of suitable size (>0.5 cm) for delivery to geochronology instruments would be straightforward and likely involve minimal processing.

However, the Spirit and InSight landing sites have a large number of small, sediment-filled craters (dubbed "hollows") dominated by accumulation of finer materials capable of aeolian transport. Material within the hollows is derived via mostly local impacts of varying size that create and eject an inventory of fines that can be transported by the wind into the source or other nearby craters where it can be sequestered. The majority of infilling occurs relatively soon after a crater forms (Golombek et al. 2014; Sweeney et al. 2018) and is related to transport of sediment produced by the impact, though prevailing winds dictate that some is transported in directions that will not contribute to infilling of the parent crater. Infilling continues over time as nearby impacts create additional inventories of fines, some of which can be transported downwind and trapped in other craters. Because wind is the dominant transport mechanism over the past ~1-2 Ga on Mars, these sediments are dominated by particles smaller than 1 cm, which are less suitable for delivery to geochronology instruments. Nevertheless, there is also a lesser component of coarser fragments (mostly cm-scale) that contributes to fill and is the result of direct emplacement of ejecta fragments during formation of nearby craters (mostly tens to hundreds of meters away). At the InSight landing site, the concentration of coarser fragments on the west side of Homestead Hollow and referred to as Rocky Field is probably an example of ejecta from a nearby crater (Grant et al. 2020; Warner et al. 2020).

Given that several prior landings on Mars (Opportunity, Spirit, InSight) have ended up in small, sediment-filled craters, a mission without active hazard avoidance capability has to ensure access to enough cm-scale fragments in the event that landing should occur within a sediment-filled crater. Information from the fill within Homestead Hollow at the *InSight* landing site and Laguna Hollow at the *Spirit* landing site suggests this should be possible. At the *InSight* landing site, the lander workspace contains ~500-600 fragments larger than 0.5 cm/m$^2$, a handful fragments >1.0 cm/m$^2$, and only ~1 fragment >2.0 cm/m$^2$, on the hollow floor (Weitz et al. 2020; Table 3). However, on the west side of the hollow ("behind" the lander), there are greater numbers (~2-4×) of cm-scale fragments in Rocky Field. An important caveat to these numbers is that the rocket motors stripped off dust and fines from the surface that likely concentrated these coarser fragments to some degree. For comparison, the fragment size distribution in Laguna Hollow in Gusev Crater (where there was no rocket motor blast) is broadly similar to that at Homestead Hollow, but there are slightly more larger fragments that increase in number toward the rim, and the fragments are slightly more elongate.

If the fragments were distributed uniformly with depth in the same density as their surface abundance, 2.7 L of material would yield ~30 fragments that are ~0.5 cm or larger in diameter (Table 3). In this case, a PlanetVac device would be appropriate for collecting the samples (though importantly, PlanetVac should be tested to ensure it can work with an indurated sediment/duricrust at the near-surface, as is seen at the *InSight* landing site). However, if the fragments represent a lag deposit and are not uniformly abundant at depth, as appears to be the case at *InSight* as a result of rocket motor blast during landing, then an arm to reach and acquire surface and near-surface samples may be more appropriate to retrieve fragments for analysis.

For Vesta, we looked at the particle size-frequency distribution in howardites, lithified regolith samples presumed to come from Vesta. A particle size-frequency distribution was reported by Pun et al. (1998) for the meteorite Kapoeta, but only for grains 0.2 cm and smaller. Nevertheless, plotting the Kapoeta grain size-frequency distribution (Fig. 7) shows the Kapoeta data at small grain sizes are a close approximation to lunar size-frequency data extrapolated to the grain sizes of interest. We may therefore be confident that the estimates developed for the lunar and martian cases would also be appropriate for the vestan regolith. Furthermore, Vesta does not have the aeolian processes that Mars does, so we would expect the variability from site to site to be more like the lunar case, and would not expect the complications that are potentially present on Mars.

## 4.2 Site Selection

To drive the engineering constraints for these studies, it was necessary to consider candidate landing sites for each body. Although these are notional landing sites based on our identified science questions, we made an effort to ensure that these sites were representative of the range of sites that the science community might desire for a mission with geochronology capabilities (Table 5).

We assessed candidate lunar sites for mission safety using *LRO* data, specifically LROC Narrow-Angle Camera (NAC) images, NAC Digital Terrain Models (DTMs), and rock abundance data derived from Diviner Lunar Radiometer Experiment measurements. NAC-scale DTMs (5 m/ px) were used to assess the topography, slopes, and roughness. Roughness was measured in terms of the Terrain Ruggedness Index (TRI), which is calculated by determining the mean elevation difference between adjacent pixels in the DTM (Riley et al. 1999; Wilson et al. 2007). TRI values (unitless) were generated using an open source Geospatial Data Abstraction Library script (GDAL 2019). Because TRI values are relative, the range of values in an area varies depending the scale of features and the range of topography. Given the range of TRI values present at our landing sites, we define values <10 as safe. We also used NAC DTMs to generate slope maps for each landing site, with low slope areas that are suitable for landing (<15°) shown in black. Finally, *LRO* Diviner data were used to assess rock abundance for each area. Diviner rock abundance (DRA) measures the cumulative areal fraction of the surface covered by rocks >1 m (Bandfield et al. 2011), with low values indicating surfaces with a low percentage of rock coverage. Suitable landing sites are those with low TRI (<10), low slopes (<15°), and low DRA values. Features of concern for the hazard assessment include impact craters, boulders, slopes, and shadows. Use of NAC images with resolution 0.5 m/px permits identification of boulders 2-3 m in diameter and larger. Using images with different illumination geometry, especially low-sun images, smaller boulders can be identified and confirmed using their elongated shadows. Diviner data can also be used to assess rock abundance over large areas and to support the *LROC* NAC data. Information derived from the *LROC* data and supported by Diviner data made it possible to place landing ellipses in areas with a suitably high probability of a lander not encountering a mission-ending hazard. With terrain-relative navigation and/or active hazard avoidance and small landing ellipse capabilities, landing sites can be selected with a very high probability of safe landing. We considered only nearside sites to maintain direct-to-earth communication capabilities.

For Objective 1, we took advantage of newly-identified potential sites to establish the chronology of lunar basin-forming impacts by measuring the radiometric age of samples directly sourced from the impact melt sheet of a pre-Imbrian lunar basin. Basin interiors are fundamentally different geologic settings than the *Apollo* sites. Basin formation is an extremely energetic process that melts large volumes of target rock. This process resets isotopic ratios in re-crystallized impact melt minerals, recording the age of basin formation (the impact melt radiometric age is technically the age of impact melt crystallization, which can lag behind basin formation by >100,000 years for thick, convective melt sheets). While some impact melt is ejected during basin formation, it is mixed with and diluted by non-melted basin ejecta and the ejecta substrate. Therefore, attempting to identify and characterize impact melt in basin ejecta has resulted in ambiguous interpretations. Conversely, the impact-melt sheet is an in-place relic of basin formation that was unaffected by the chaotic ejecta process. Directly sampling the impact-melt sheet would provide the most reliable record of the basin's age. In most of the nearside basins, volcanic activity flooded the basins, covering their impact-melt sheets. But in several locations, younger impact craters punched through this basaltic veneer to expose the original impact melt sheet, as revealed by their geochemical signatures from orbit.

The Crisium basin has a model age of ≥3.94 Ga (van der Bogert et al., 2017) and ≤4.07 Ga (Orgel et al. 2018) as determined by crater size-frequency distribution measurements, but is significantly overprinted with Imbrium secondaries. An *in situ* dating precision of 0.20 Ga (2σ), may not be enough to distinguish the age of Crisium from Imbrium, but an age constraint on Crisium might increase the number of basins known to have formed in a cataclysm, including Crisium and Humboltianum (which sits stratigraphically between Crisum and Imbrium (Fassett et al. 2012). A greater number of basins (~12) has CSFD-derived model ages suggesting that they formed between Nectaris and Imbrium. If the age difference between the Nectaris and Imbrium is negligible, then a greater number of basins had to form in a short time strengthening a cataclysm interpretation. But if the age difference between Nectaris and Imbrium were measurable, then those twelve basins would have had a longer time window in which to form. As yet, no returned samples have been definitively linked to the Nectaris basin impact-melt sheet. Consequently, the Nectaris basin has been interpreted to be as young as 3.85 Ga or as old as 4.17 Ga (James 1981; Neukum & Ivanov 1994; Stöffler et al. 2006; Fischer-Gödde & Becker 2012; Orgel et al. 2018), though the younger interpretation is inconsistent with geological mapping and stratigraphy (Fassett et al. 2012). Differences between the two suggested ages may be resolvable within the precision assumed in this study. If the age were 3.8±0.2 Ga, we would appear to confirm the younger option; if it were 4.1±0.2 Ga, we would confirm the older age. However, on the face of it, a single age determination having a 200 Myr uncertainty may be insufficient to confidently distinguish the age of these early basins from the Apollo sample-derived age of Imbrium. Therefore, the aforementioned strategies of using two independent techniques on multiple samples must be more rigorously developed to understand how best to address this objective.

In Crisium, several candidate impact-melt exposures were identified by Spudis and Sliz (2017). In examining these candidate sites, Runyon et al. (2020) investigated Yerkes crater (36 km, 14.6° N, 51.7° E) as a promising exposure of the Crisium impact-melt sheet. Yerkes predates the most recent mare flows, but the rim and central peaks of Yerkes have not been fully buried by those flows. The rim and central peak structures exhibit a noritic mineralogical signature, distinguishing Crisium impact melt from the mare basalts (Runyon et al. 2020). Similar compositional patterns are observed at other craters within central Crisium, including Peirce (18.8 km, 18.26° N, 53.35° E), Picard (22.3 km, 14.6° N, 54.7° E), Lick (31.6 km, 12.4° N, 52.8° E), and several smaller craters, where noritic impact melt is exposed from beneath the mare fill. For this study, we further characterized Peirce as a potential landing site (Figs. 8, 9). Based on crater degradation and embayment relationships apparent in LROC WAC imagery and LOLA topography, Peirce appears to postdate the most recent mare basalt emplacements. The floor of Peirce exhibits a distinctive noritic character in $M^3$ mafic abundance and pyroxene composition parameters generated using techniques developed and validated by Moriarty III and Pieters (2016). This composition is similar to the rim and central peak of Yerkes, indicating that Peirce also exposed the pre-mare Crisium impact-melt sheet.

As at Crisium, the melt sheet of Nectaris has been mostly obscured by post-basin mare resurfacing. However, Rosse crater (11.4 km, 17.9° S, 35.0° E) in central Nectaris appears to have excavated through the mare basalts, exposing noritic impact melt material from the melt sheet (Figs. 10, 11). In Rosse, impact melt material is abundant and mineralogically and spectroscopically distinct from local mare basalts, allowing unambiguous *in situ* identification. While each of these features has exposed the underlying basin impact-melt sheet, landing sites with access to material from crater central peaks may be good targets for obtaining the age of the basin itself, rather than the reset age of the younger crater (Young et al. 2013). Furthermore, identifying the composition of Nectaris impact melt may make it possible to definitively identify basin-specific impact-melt samples in the Apollo sample collection.

For Objective 2, we considered two sites: P60 basalt and Le Monnier crater to establish the age of a very young lunar basalt to correlate CSFD measurements with a radiometric crystallization age. The P60 basalt unit, located just south of the Aristarchus plateau (approximately 21°N, 40°W), has a CSFD age as young as ~1 Ga (Hiesinger et al. 2003, 2011; Stadermann et al. 2018, which makes it the youngest observed extensive mare basalt unit on the lunar surface. The P60 unit is a prime target for geochronology studies because dating a young lunar surface would help to anchor the young end of the chronology function (Jawin et al. 2019). Accordingly, several robotic missions have been proposed to go to this uniquely young basalt (Carson et al. 2016; Draper et al. 2019). We used $M^3$ data to assess the mineralogical diversity of the P60 area using mafic mineral abundance and pyroxene composition (Moriarty III & Pieters 2016). The P60 mare emplacement exhibits a distinctly basaltic mineralogy based on strong, relatively long-wavelength spectral absorption bands (Fig. 12). Most of the P60 area is relatively flat, has low slopes, and low TRI values. However, owing to its young age, the regolith is relatively immature, resulting in areas of higher DRA than older mare surfaces. Figure 13 outlines two potential landing areas that have low slopes and low TRI, as well as few measurable boulders.

Le Monnier is a 57-km diameter impact crater on the western rim of Serenitatis basin, at 26.86°N, 30.08° E. The eastern rim of Le Monnier is absent and the crater is infilled with mare basalt, forming a bay off of Mare Serenitatis. The basalt pond that covers the surface of Le Monnier has a model age of 2.4 Ga, based on crater counts (Hiesinger et al. 2011). The young age of the basalt in Le Monnier makes it a particularly interesting target. Geologic context and mineralogic diversity for the Le Monnier region (Fig. 14) show that the crater floor also has a distinctly basaltic mineralogy, based on strong, relatively long-wavelength spectral absorption bands in $M^3$ data. Landing site assessment shows that there are many possible safe landing ellipses in the smooth, flat interior or Le Monnier. This makes it an ideal candidate for a landed mission (Fig. 15) and a complement to the P60 basalt or a comparably young basalt, such as the target of *Chang'E 5* sample return, northeast of Mons Rümker (Zhao et al. 2017).

Our Mars-specific study aims included Objective 3 (establish the age of a well-exposed Hesperian martian lava terrain to correlate crater density with crystallization age) and Objective 4 (establish the epoch of martian habitability by measuring the radiometric age of Noachian clay-bearing stratigraphies). We initially identified sites that would enable access to both lithologies of interest, taking advantage of the significant engineering and scientific research expended on potential landing sites for previous, current and future landed missions to maximize confidence in accessibility and interpretations. Examples include the broad lava plains of Syrtis Major, where lavas are exposed at the Nili Fossae Trough and to the south of the NE Syrtis candidate landing sites for *Curiosity* and *Perseverance*. The Nili Fossae Trough (Fig. 16) provides access to representative sections of widely distributed units, including Noachian units with clay minerals and Hesperian lavas. The Northeast Syrtis region also provides access to representative sections of widely distributed units including clays, carbonates, sulfates, and lavas (Fig. 17).

Extensive characterization studies in this area provide the geologic context with which to interpret geochronology dating (Mustard et al. 2007; Ehlmann & Mustard 2012; Bramble et al. 2017; Scheller & Ehlmann 2020).

The addition of mobility to a geochronology mission to either of these sites could enable access to lavas and clay-bearing strata within short distances of the landing site (~5 km for Nili Fossae Trough, ~20-30 km for NE Syrtis), and could enable achieving both Mars 1 and Mars 2 objectives. Both sites would also be suitable for a fixed lander, though a fixed lander would enable access to only one type of terrain and solve only one objective. Additional examples exist across Mars for one or both of these objectives, such as the landing sites considered as candidates for the *Curiosity and Perseverance* missions (Grant et al. 2011; Grant et al. 2018). Other widely exposed lava plains occur as candidate landing sites (e.g., Hesperia Planum), though these sites would require considerable additional analysis to certify. Mawrth Vallis (Fig. 18) represents a widespread, ancient clay-bearing sequence of rocks that has also been well-characterized for suitable landing (Poulet et al. 2020). Other examples include sedimentary/hydrothermal materials with mineralization conducive to age-dating such as the jarosite-sulfate sediments at NE Syrtis Major (Quinn & Ehlmann 2019), jarosite-bearing weathering sequences or sediments both inside and outside of Valles Marineris (Milliken & Bish 2010; Weitz et al. 2015), and jarosite and alunite within clay-bearing sedimentary deposits in Columbus or Cross crater paleolakes (Wray et al. 2011; Ehlmann et al. 2016b).

Expanding the absolute chronological framework of an asteroid within the precision of *in situ* dating would be most impactful by establishing the radiometric ages of samples with well-established provenance. This constraint limits landing sites to well-studied bodies with specific geologic epochs, such as Vesta. Key stratigraphic craters and contiguous geologic terrains would yield age determinations that constrain the body's geologic timescale. The most prominent impact structures on Vesta, and those that pin its stratigraphy, are the Veneneia, Rheasilvia and Marcia craters (Fig. 19). Ideal landing sites would allow sampling of all three key impact structures, as well as some of the youngest impact melt deposits. Here we assume that sampling would be similar to a mature lunar soil, in terms of the number of fragments available and the geologic diversity of the fragments. The best-resolution images of Vesta from the Dawn mission's low altitude mapping orbit are 70 meters per pixel, too coarse to identify hazards at the lander scale, meaning a mapping survey would be needed to identify suitable landing locations.

The first site considered on Vesta was the Rheasilvia central peak. Deep-seated material brought to the surface would yield information about internal structure and composition (e.g., potential mantle material). The flat, high-standing plateau of the central peak means resurfacing should be minimal, yielding a good location to derive the age of Rheasilvia. Because published ages range from 1-2 Ga to 3.4±0.1 Ga, this is a prime target for *in situ* geochronology, assuming the samples can be tied to Rheasilvia through chemical or other means. The second site we studied was Marcia crater, the youngest large crater on Vesta (Williams et al. 2014a). Dark Veneneia material may be exposed in Marcia, as well as superposed bright Rheasilvia material and Marcia ejecta material. The age of Marcia is not as contentious as the age of Rheasilvia or Veneneia, but a well-studied suite of samples at this one site could potentially provide as many as three tie-points for calibrating the chronology of Vesta, and by extension the entire middle portion of the main asteroid belt. A mission to such a complex site would require analyzing more samples at the site, nominally 10 for each of the three target lithologies.

## 5. Geochronology Mission Implementation

All Geochronology mission studies were conducted at Concept Maturity Level (CML) 4: an implementation concept at the subsystem level, as well as science traceability, key technologies, heritage, risks and mitigations. In addition, the team developed cost models. Some CML 5 aspects were also accomplished, including requirements traceability and notional schedules to the subsystem level. The studies assumed that each mission would be a Class B, PI-led mission, consistent with a New Frontiers Announcements of Opportunity.

Functionally, a mission to accomplish these science objectives must be a landed mission with access to surface samples. The mission must be capable of collecting, triaging, and analyzing ten 0.5-2 cm samples of each target lithology. The mission must also be capable of remotely sensing the lander workspace and the landing site to provide sample context and create spatially contiguous maps of the landing site for orbital context. The outcome of some important trades is presented in Table 6. All versions of the Geochronology mission in this study closed using a single lander (Fig. 20), with the capability to hop to a second site implemented for the Vesta design. The requirement for *in situ* analysis necessitated a lander design and the goal to stay within the anticipated New Frontiers 6 cost cap precluded mobility solutions on the Moon and Mars. Detailed engineering designs for the lander concept studies may be found in the full report (Cohen et al. 2020).

The notional payload would provide a complementary and robust approach to resolving the science goals using technologies that would be ready for a New Frontiers 6 Announcement of Opportunity. A common payload would be implemented for all destinations, with minor changes to adapt to the Martian environment. The selected hardware is meant to be representative of a payload that could meet the science requirements and be accommodated on a New Frontiers-class mission, rather than endorsing an instrument or specific payload configuration. The payload mass sums to approximately 180 kg, which includes 30% margin over the current best estimates. Peak power draws would come from the laser operations during CDEX, KArLE, and ICP-MS analysis. Downlink needs would be driven by the imagers and imaging spectrometer.

The geochronology landers would use heritage hardware and structural designs that are well within the state of the art. All would be able to launch on existing launch vehicles, such as the Falcon 9 Heavy. All landers would use solar arrays and batteries to support power and communication needs for the payload, lasting one terrestrial year on the surface of each destination. Destination-specific factors in the spacecraft bus designs include the thermal solution, engines and propellant volume, and entry, descent, and landing (EDL) details. The Deep Space Network would be used as the primary means for all communications during the flight operations, checkout, commissioning, and surface operations phases. Each lander would use direct-to-Earth communication links to forward commands to and receive science and housekeeping data. After landing, a several-day lander checkout would be performed, followed by instrument commissioning and initial data collection. Normal surface operations would occur during the daylight portion of the lunar or martian day, when the lander would perform sample analysis and site imaging while recharging the lander batteries and communicating with Earth. Then it would enter a low-power state for the night. On Vesta, the available power would allow the sample analysis sequence to occur uninterrupted and would not need to be paused during the Vesta night.

Remote-sensing instruments would be mast-mounted with full 360° azimuth coverage and elevation from lander to horizon. Sample analysis instruments would be arranged in an arc around the Sample Manipulation Arm (SMA) (Fig. 21). The team developed a reference sample analysis process through which the samples would be acquired, analyzed, prioritized, and sequentially introduced to each instrument (Fig. 22). The reference sequence would be repeated for each sample until the required number of samples were completed at each site. The sequence may be paused at multiple points for additional data analysis, troubleshooting, or to manage payload power. PlanetVac would pneumatically gather surface samples, transfer, and sieve them to the 0.5-2 cm diameter required for analysis. The sieved samples would be gravity-fed into the triage station for identification and prioritization using data from the spectrometer and Microimager. The operational time required for characterization and prioritization would depend on the process used. It should be possible to develop onboard algorithms to characterize at least some samples using preselected characteristics known from laboratory analysis, but some scientific ground-in-the-loop characterization may also be required. After analysis of the initial triage data, prioritized samples would be selected for further analysis. The SMA would move samples from the triage station to the other instruments. CDEX requires that the rock sample present a flat, polished surface for analysis, so the SMA would first present a sample to a grinding station. At the grinding station, the rock would be ground to millimeters of depth and polished to a 10-micron finish using a heritage design Rock Abrasion Tool (RAT), previously flown on the *Mars Exploration Rovers* (*MER*) *Spirit* and *Opportunity*. After polishing, the sample would be presented for imaging of the ground surface and then to the CDEX aperture, where CDEX would perform its analysis by rastering the flat, finished surface of the sample over the instrument entrance aperture. Typically, CDEX would measure 100-300 locations on a sample in a raster pattern, ablating 5-μm diameter laser spots, thus sampling a range of different minerals for Rb-Sr isotope ratio measurements. The ionization lasers could be turned off to map elemental chemistry.

Upon completion of CDEX analysis, the SMA would drop the sample into the KArLE/ICP-MS carousel inlet. In this study, KArLE and ICP-MS would share a laser ablation unit, a mass spectrometer, and an internal sample-handling carousel, which is based on the Sample Analysis at Mars (SAM) sample carousel on the Mars Science Laboratory (MSL) mission. Neither KArLE nor ICP-MS require samples to have a flat, polished surface, but can accept either a natural sample or a polished sample. Therefore, if the grinding or imaging station fail, KArLE and ICP-MS could still analyze the sample. However, samples could not be retrieved from the KArLE carousel after analysis, so this step occurs last in the operations. The sample would be dropped into an individual cup in the carousel; the carousel would rotate the sample cup to the analysis station and elevate the cup to form a seal. The sample would be laser ablated to measure elemental chemistry (including parent K) using LIBS. Material liberated by laser ablation would be let into a mass spectrometer either by static expansion and electron impact ionization (neutral species including Ar for KArLE analysis) or via a plasma torch (ionized species including trace elements for ICP-MS).

While KArLE and ICP-MS analyze the rock sample, the SMA could return to the triage station for the next rock sample. After collection and triage, the total analysis cycle for one rock sample would be performed within 24 hours. This cycle would be repeated 20 times at the Mars and lunar sites and 10 times at each of the Vesta landing sites. The triage process could be repeated multiple times in the surface operations timeline to obtain a fresh set of samples. In parallel with sample science, the Panoramic Imagers and UCIS spectrometer would obtain 360° coverage of the landing area from lander to horizon. Because of lighting constraints, the collection cadence of the context images would be specific to each destination.

A total of 20 samples would be analyzed in all concepts, so the data generated by the geochronology instruments would be the same for all destinations. On Vesta, where there would be two science sites, two context imaging datasets would be acquired and there would be additional triage processes planned. Data summaries for the Lunar, Vesta, and Mars concepts are shown in Table 7.

The team used parametric cost modeling to estimate costs for instruments and landing systems and wrap factors for the other spacecraft development elements, and added 50% unallocated margin to the total. The total costs of all mission concepts are within family of a New Frontiers mission classification. The Phase B-D durations, milestone reviews and general schedule flow of the Geochronology missions in this study are well within family of previous New Frontiers missions though they would vary slightly in their launch readiness dates and constraints and Phase E elements and durations. Detailed costing methodology and cost breakdowns can be found in the Full Report (Cohen et al. 2020).

## 6. Summary and Recommendations

Table 8 summarizes all the studied architectural options and how they would meet Geochronology science goals and mission drivers. New Frontiers-class landers could carry full payloads for ~1 year of operations to a single site on the Moon and Mars. The Vesta architecture option would meet sample science objectives at multiple sites within a New Frontiers cost cap. Although study constraints on the cost and payload mass preclude significant mobility on Mars and the Moon, whether by hopper or rover, sites may exist where multiple objectives could be met by analyzing more samples or complementing sample-return efforts. Reductions to the payload, such as descoping one radiometric system or limiting the sample characterization suite, could be considered to reduce payload mass, potentially mitigated by judiciously choosing focused science questions and sites where they could be answered.

These studies make it clear that feasible New Frontiers-class missions could carry a capable instrument payload to conduct *in situ* dating with the precision to answer community-identified Geochronology science goals. Recent NASA investments in *in situ* dating instruments have resulted in payload options that will be ready to infuse into these or other missions in the next decade. Missions that fit within the New Frontiers-class and include dating by multiple corroborating methods and extensive characterization to increase confidence in results are possible. Additionally, new remote sensing, geologic mapping, and site evaluation efforts have expanded the locations where safe landing sites can access lithologies of interest.

Although mobility options were too expensive to fit into the New Frontiers cost cap in this study, compelling cases can be made for targeted single-site landers on the Moon and Mars to address individual Science Objectives. Such missions would also enable a broad suite of geologic investigations, such as basic geologic characterization; geomorphologic analysis; establishing ground truth for remote sensing analyses; analyses of major, minor, trace, and volatile elements; atmospheric and other long-lived environmental monitoring; organic molecule analyses; and soil and geotechnical properties.

It is conceivable the current NASA Commercial Lunar Payload Services (CLPS) program will survive and grow into the next decade. If so, the larger landers could mature into options that could host part or even all of the Geochronology payload. The growth areas needed include total payload mass capability, broader landing site capability, higher reliability, lower risk, and the addition of packages that provide power and thermal solutions (such as the extended operations package capability of the Lunar Environment Monitoring Station; Benna et al.,. 2020) for the lunar night. Continuing the CLPS program into the next decade would increase the likelihood that lower-cost communications relays would become available. If the goals and promises of the CLPS program also result in lower payload delivery cost, multiple lunar Geochronology sites could be investigated on the Moon, potentially at lower cost.

Additional technology maturation investments beneficial to this class of mission may include improving the sensitivity of geochronology instruments to enhance age measurement precision, developing end-to-end sample acquisition and handling to feed samples to multiple instruments (e.g., *Curiosity*, *Europa Lander*; Anderson et al. 2012; Malespin et al. 2020), and expanding flight system technologies to enable spacecraft operations that increase science return (e.g., high-performance computing chips and boards capable of processing terrain navigation and hazard avoidance algorithms, more efficient batteries, thermal technologies that enable night survival, communications throughput).

*In situ* geochronology is a feasible way to address community-identified science goals at the Moon, Mars, and Vesta. The Geochronology team advocates that NASA include opportunities in the New Frontiers missions list for answering these compelling science questions, with the implementation flexibility to meet them by sample return or *in situ* dating.

Table 1: Summary of Geochronology Science Goals and Objectives for the Moon, Mars, and Vesta and traceability to candidate measurements and notional payload.

<table>
<tr><th>Science Goal</th><th>Science Objective</th><th>Traceability</th></tr>
<tr><td rowspan="2">Goal A: Determine the chronology of basin-forming impacts to constrain the time period of heavy bombardment in the inner Solar System</td><td>Objective 1: Establish the chronology of basin-forming impacts by measuring the radiometric age of samples directly sourced from the impact melt sheet of a pre-Imbrian lunar basin.</td><td>SCEM 1a. Test the cataclysm hypothesis by determining the spacing in time of the creation of lunar basins</td></tr>
<tr><td>Objective 5: Establish the radiometric ages of vestan samples with well-established provenance.</td><td>SBAG 1.2.2. Determine the timing of events in the early Solar System, and 1.2.3. Use the distribution of compositions and ages of small bodies in the Solar System to make testable predictions about observable parameters in forming planetary systems.</td></tr>
<tr><td rowspan="2">Goal B: Reduce the uncertainty for inner Solar System chronology in the “middle ages” (1-3 Ga) to improve models for planetary evolution</td><td>Objective 2: Establish the age of a very young lunar basalt to correlate crater size-frequency distributions with crystallization ages.</td><td>SCEM 5b. Determine the age of the youngest and oldest mare basalts; and 5d. Determine the flux of lunar volcanism and its evolution through space and time.</td></tr>
<tr><td>Objective 3: Establish the age of a well-exposed Hesperian martian lava terrain to correlate crater size-frequency distributions with crystallization ages.</td><td>iMost 1.5 Determine the petrogenesis of martian igneous rocks in time and space; and iMost 5.0 Reconstruct the history of Mars as a planet</td></tr>
<tr><td>Goal C: Establish the history of habitability across the Solar System.</td><td>Objective 4: Establish the epoch of martian habitability by measuring the radiometric age of Noachian clay-bearing stratigraphies.</td><td>iMost 3.0 Determine the evolutionary timeline of Mars; and 4.0 Constrain the inventory of martian volatiles as a function of geologic time and determine the ways in which these volatiles have interacted with Mars as a geologic system.</td></tr>
</table>

<table>
<tr><th>Measurement Goals</th><th>Measurement Requirements</th><th>Payload Element[2]</th><th>Functional Requirements</th></tr>
<tr><td rowspan="2">Measure the age of the desired lithology with precision ±200 Myr</td><td>Use Rb-Sr radiometric chronology to directly measure the age of samples derived from the target lithology</td><td>CDEX</td><td rowspan="7">Collect, triage, and analyze samples of each target lithology<br><br>Remotely sense the lander workspace to provide sample context at the landing site to create spatially contiguous maps</td></tr>
<tr><td>Use K-Ar radiometric chronology to directly measure the age of samples derived from the target lithology</td><td>KArLE</td></tr>
<tr><td rowspan="4">Contextualize the desired lithology using petrology, mineralogy, and/or elemental chemistry</td><td>Measure major- and trace-element geochemistry of the samples to establish parentage and evolution of lithologies</td><td>ICPMS, CDEX, KArLE</td></tr>
<tr><td>Identify mineralogy by mapping abundances of olivine, pyroxene, oxides, plagioclase, and aqueous alteration minerals</td><td>UCIS</td></tr>
<tr><td>Image samples at the microscale to determine grain size and petrology</td><td>Microimager</td></tr>
<tr><td>Determine the composition of the surface unit to place the lithologies into a regional and global context</td><td rowspan="2">Panoramic Imager, UCIS</td></tr>
<tr><td>Relate the measured lithology age to crater counting of the lithology’s terrain</td><td>Determine the geology of the landing site and map discrete lithologic units to relate them to maps and crater counts determined from remote sensing</td></tr>
</table>

Table 2: Notional Geochronology payload heritage and technology readiness levels.

| Payload Element | Provider | Flight heritage | Maturation path (NASA investments)[1] | Current Status | Projected TRL 2023 | | |
|---|---|---|---|---|---|---|---|
| | | | | | Moon | Mars | Vesta |
| CDEX | SwRI | -- | PIDDP (2013), MatISSE (2017), Proposed to Mars 2020 and Discovery 2015 | Performance demonstrated in functional breadboard | 6 | 6 | 6 |
| KArLE | GSFC | GSFC mass spectrometers including SAM, LADEE, MOMA-MS, LIBS from ChemCam, carousel from SAM | PIDDP (2014), DALI (2018), proposed to Mars 2020 | Performance demonstrated in functional breadboard | 6 | 6 | 6 |
| ICPMS | University of Maryland | GSFC mass spectrometers including SAM, LADEE, MOMA-MS | PICASSO (2018), SBIR/STTR; additional MatISSE or DALI to achieve TRL 6 | Proof of concept demonstrated | 4 | 4 | 4 |
| UCIS | JPL | Moon Mineralogy Mapper, others | MatISSE (2012) for Mars and Vesta, DALI (2018) for Moon | Performance demonstrated in functional breadboard | 6 | 6 | 6 |
| Panoramic and Microimager | MSSS | MAHLI, WATSON, ECAMS (MSL and OSIRIS-REx) | CLPS (2022) | Flown in relevant environments | 9 | 9 | 9 |
| PlanetVac and other hardware | Honeybee Robotics | -- | MMX, CLPS | Performance demonstrated in functional breadboard | 9 | 6 | 9 |

[1]Planetary Instrument Concepts for the Advancement of Solar System Observations (PICASSO); Maturation of Instruments for Solar System Exploration (MatISSE); Development and Advancement of Lunar Instrumentation (DALI); Small Business Innovation Research (SBIR) / Small Business Technology Transfer (STTR); Commercial Lunar Payload Services (CLPS); Martian Moons eXplorer (MMX)

Table 3: Rock size-frequency calculations for the Moon, Mars, and Vesta.

| | Diameter (cm) | Surface density ($m^{-2}$) | Volume density ($L^{-1}$) [1] | Volume required for 30 rocks (L) |
|---|---|---|---|---|
| Moon (boulders)[2] | 0.5 | 9.89E+03 | 9.83E+02 | 2.58E-02 |
| | 1 | 2.88E+03 | 1.55E+02 | |
| | 2 | 8.39E+02 | 2.43E+01 | |
| Moon (cores)[3] | 0.5 | − | 4.42E+01 | 6.17E-01 |
| | 1 | − | 3.93E+00 | |
| | 2 | − | 5.00E-01 | |
| Mars (hollows)[4] | 0.5 | 5.00E+02 | 1.12E+01 | 2.68E+00 |
| | 1 | 1.00E+01 | 3.16E-02 | |
| | 2 | 1.00E+00 | 1.00E-03 | |
| Vesta[5] | 2.00E-06 | − | 1.78E+13 | Similar to lunar |
| | 2.00E-05 | − | 7.07E+11 | |
| | 2.00E-01 | − | 9.71E+03 | |

[1]calculated using (surface density)$^{3/2}$
[2]Shoemaker and Morris (1970)
[3]median values from 21 core samples
[4]order-of-magnitude visual estimates from InSight images
[5]Pun et al. (1988)

Table 4: Rock size-frequency counts for lunar cores.

| Sample | Volume (cm$^3$) * | Maturity | Rock Count | | |
|---|---|---|---|---|---|
| | | | 0.5-1.0 cm | 1.0-1.5 cm | 1.5-2.0 cm |
| 10004 | 41.8 | | 6 | 0 | 0 |
| 10005 | 29.9 | | 5 | 0 | 0 |
| 12026 | 60.6 | | 3 | 0 | 0 |
| 12027 | 53.4 | Submature-mature | 5 | 0 | 0 |
| 12025 & 12028 | 128.8 | | 5 | 1 | 0 |
| 14210 & 14211 | 118.1 | Mature | 16 | 4 | 0 |
| 14220 | 51.8 | Mature | 10 | 0 | 0 |
| 14230 | 39.3 | Submature | 3 | 0 | 0 |
| 15009 | 377 | Submature | 43 | 8 | 1 |
| 15007 & 15008 | 711.3 | Submature | 17 | 3 | 0 |
| 15011 & 15010 | 841.9 | Maturity decreases with depth | 38 | 8 | 6 |
| 60010 & 60009 | 738.9 | | 32 | 9 | 3 |
| 60014 & 60013 | 792.9 | | 5 | 0 | 1 |
| 64002 & 64001 | 824.4 | | 12 | 3 | 0 |
| 68002 & 68001 | 782.9 | | 59 | 13 | 0 |
| 70012 | 231.2 | | 4 | 1 | 1 |
| 76001 | 433.5 | | 1 | 0 | 0 |
| 73002 | 276.5 | | 7 | 2 | 2 |
| 74002 & 74001 | 857 | | 1 | 3 | 0 |
| 79002 & 79001 | 644.7 | | 16 | 8 | 6 |

*Calculated using recovered sample core length, assuming 100% filling of the core tube.

Table 5. Summary of potential landing sites considered for this study.

| Body | Science Objective | Site | Location | Characteristics |
|---|---|---|---|---|
| Moon | Objective 1: Establish the chronology of basin-forming impacts by measuring the radiometric age of samples directly sourced from the impact melt sheet of a pre-Imbrian lunar basin. | Peirce Crater | 18.26°N, 53.35°E | 18-km diameter crater, excavates noritic Crisium impact-melt floor. Several-km landing ellipses exist. |
| | | Rosse Crater | 17.9°S, 35.0°E | 11-km diameter crater, excavates noritic Nectaris impact-melt floor. Several-km landing ellipses exist. |
| | Objective 2: Establish the age of a very young lunar basalt to correlate crater size-frequency distributions with crystallization ages. | P60 basalt | 21°N, 40°W | Multiple flow units as young as ~1 Ga. Widely accessible, safe landing sites exist. |
| | | Le Monnier Crater | 26.86°N, 30.08 W | 57-km diameter crater embayed by Mare Serenitatis; basalt age ~2.4 Ga. Several-km landing ellipses exist. |
| Mars | Objective 3: Establish the age of a well-exposed Hesperian martian lava terrain to correlate crater size-frequency distributions with crystallization ages.<br><br>Objective 4: Establish the epoch of martian habitability by measuring the radiometric age of Noachian clay-bearing stratigraphies. | Nili Fossae Trough | 74.481°E, 21.0108°N | Provides access to representative sections of widely distributed units, including Noachian units with clay minerals and Hesperian lavas. Ellipse was proposed and vetted for Mars 2020 Landing Site selection. |
| | | NE Syrtis | 77.0767°E, 17.8034°N | Access to a broad range of Noachian and Hesperian materials: clays, carbonates, sulfates, lavas. Landing ellipses not fully vetted for Mars 2020 |
| | | Mawrth Vallis | 21.1343°W, 24.5537°N | Access to representative sections of widely distributed units including Noachian clay-bearing stratigraphies and Hesperian dark mantling materials. Ellipse was proposed and vetted for Mars 2020 Landing Site selection. |
| Vesta | Objective 5: Establish the radiometric ages of vestan samples with well-established provenance. | Rheasilvia Basin | 71.95°S, 86.30°E | Flat, high-standing plateau at basin central peak enables access to basin material. Current best image resolution ~70 m/px. |
| | | Marcia Crater | 15°N, 180°E | Marcia is a key stratigraphic marker, location sited among a variety of geologic units spanning geologic history. Current best image resolution ~70 m/px. |

Table 6. Important trade space outcomes for a potential Geochronology lander.

| Trade | Option | Results |
|---|---|---|
| Mobility | Hopping | Hopping feasible on Vesta but, hopping drove propulsion system to be large and costly on the Moon |
| | Rover | Rover-based mobility out of cost cap for Mars |
| | None | Increased the number of samples to be studied at a single site to compensate when mobility could not be achieved |
| Launch Vehicle | Falcon 9 Heavy Recovery 7,049 kg | Both the Atlas 551 and Falcon 9 Heavy Recovery would be viable options for mission launch mass and fairing size. Due to expected availability issues with the Atlas 551, the team selected the Falcon Heavy Recovery. |
| | Atlas 551 6,330 kg | |
| | Delta IV Heavy 10,566 kg | |
| | ULA Future Vulcan w/Centaur 8,299 kg | |
| Descent Propulsion | Chemical | Chemical would be lowest cost and best attitude control and landing velocity |
| | Solid | Solid would be required if chemical needs large burn times to reduce finite burn losses, but selected chemical design would not have a finite burn time issue |
| | Mix | A mix of propulsion types would add complexity and cost. Most efficient if hopping. |
| Surface Power | Fuel cells | Fuel cells cannot survive the lunar day and night cycle for the mission duration |
| | Next-Generation Radiothermal Isotope Generator (NGRTG) | NGRTG would add costs, regulatory and thermal issues |
| | Solar with batteries | Solar arrays with batteries were selected for all missions |
| Land Safely | Probabilistic approach | Use knowledge of landing site and landing ellipse to define site with low risk |
| | Terrain Relative Navigation | Use pre-existing maps of landing area to identify hazards and define flight path |
| | Terrain Relative Navigation with Hazard Avoidance | Adds ability to use ACS thrusters to avoid hazards during descent on defined flight path |

Table 7. Payload.

| Payload Element | Mass (kg)[1] | Peak Power (W)[1] | Data Generation (Mbit)[2] |
|---|---|---|---|
| CDEX | | | |
| CDEX instrument | 71.5 | 182 | 22400 |
| Grinding station | 7.41 | 26 | N/A |
| Postgrind Imager | 0.78 | 9.1 | 1500 |
| Sample Manipulation Arm | 13 | 26 | 1600 |
| *Vacuum chamber & pump (Mars only)* | *6.67* | *15.6* | *N/A* |
| KArLE | | | |
| KArLE Instrument[3] | 29.77 | 130 | 21220 |
| *Vacuum pump (Mars only)* | *0.169* | *15.6* | *N/A* |
| ICPMS3 | 12.4 | 132.6 | 38 |
| UCIS (Including DPU) | 6.5 | 39 | 11268 |
| Panoramic Imagers (total for 2) | 1.508 | 19.2 | 1454 |
| Microimager | 1.43 | 9.75 | 180 |
| Imaging DVR[4] | 1.43 | 0 | N/A |
| Sample acquisition and triage | | | |
| PlanetVac | 20.8 | 41.6 | 30 |
| Triage station | 3.25 | 7.8 | N/A |
| Electronics box | 2.99 | 29.6 | |
| Totals | 172.7/*179.5* | | 59690 |

[1]Includes 30% margin on current best estimates

[2]Total for 20 samples and one location (so Vesta would double the imaging data). Data values for imagers, CDEX, and UCIS include onboard compression.

[3]KArLE and ICPMS would share laser ablation and mass spectrometer subsystems, these masses are not double-booked

[4]A single 4-port Digital Video Recorder (DVR) would serve the Postgrind Imager, Panoramic Imagers, and Microimager

Table 8: A summary of architecture options in this study and assessment against Science Goals and Objectives shows there are multiple compelling mission options in the New Frontiers family. The color code shows how well each architecture option could carry a full payload, conduct measurements at two sites on each body, and meet a New Frontiers cost cap (green = substantively met; yellow = partially met; orange = did not meet).

| **Target** | **Science Goal** | **Payload** | **Multiple Sites** | **Cost Class** |
|---|---|---|---|---|
| Moon | Determine the chronology of basin-forming impacts | Full | Single lander | New Frontiers |
| | Constrain uncertainty in lunar chronology from 1-3 Ga | Full | Single lander | New Frontiers |
| | Both | Reduced | Hopper 100s of km | Flagship |
| Mars | Validate crater-counting ages on Mars | Full | Single lander | New Frontiers |
| | Bound the epoch of habitability | Full | Single lander | New Frontiers |
| | Both | Reduced | Rover 10s of km | Flagship |
| Vesta | Establish vestan chronology | Full | Hopper 100s of km | New Frontiers |

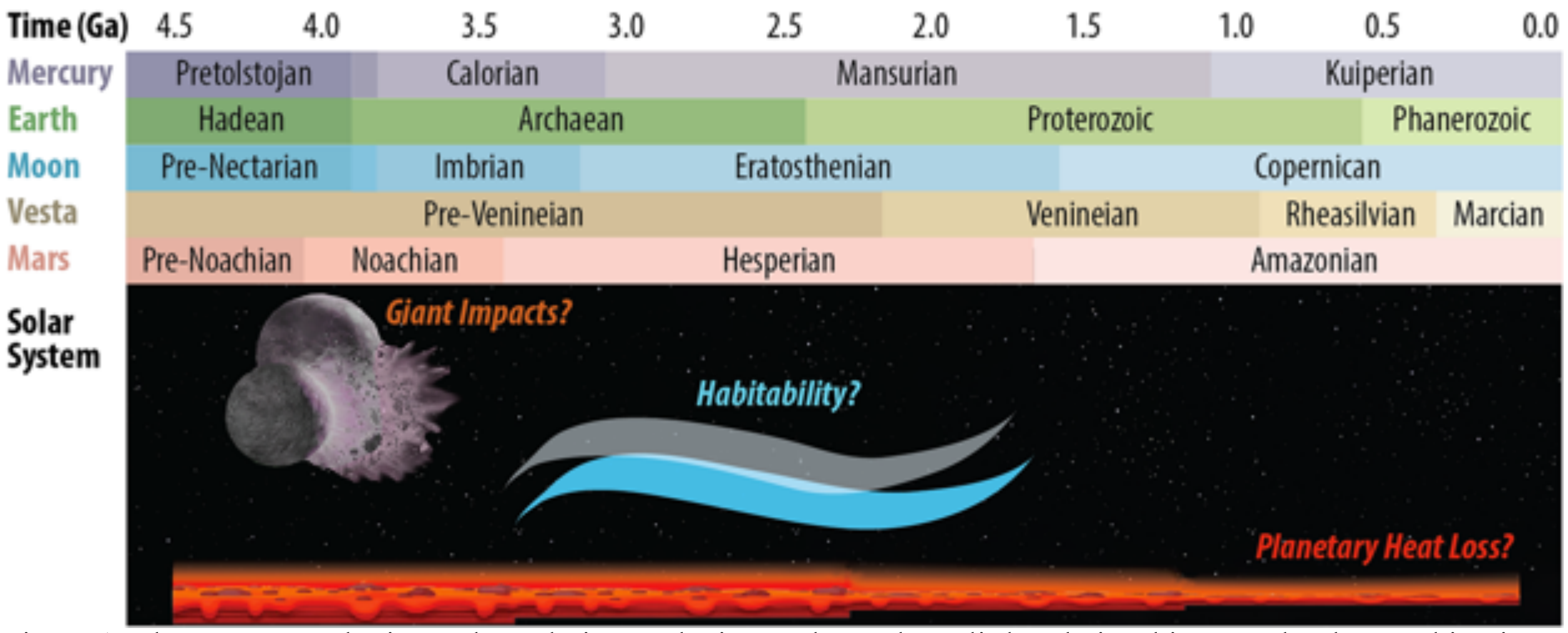

Figure 1: The current geologic age boundaries on the inner planets have little relationship to each other, making it challenging to interpret geologic evolution within a solar system context. Uncertainty in the relationship between each geologic era could be resolved using *in situ* dating.

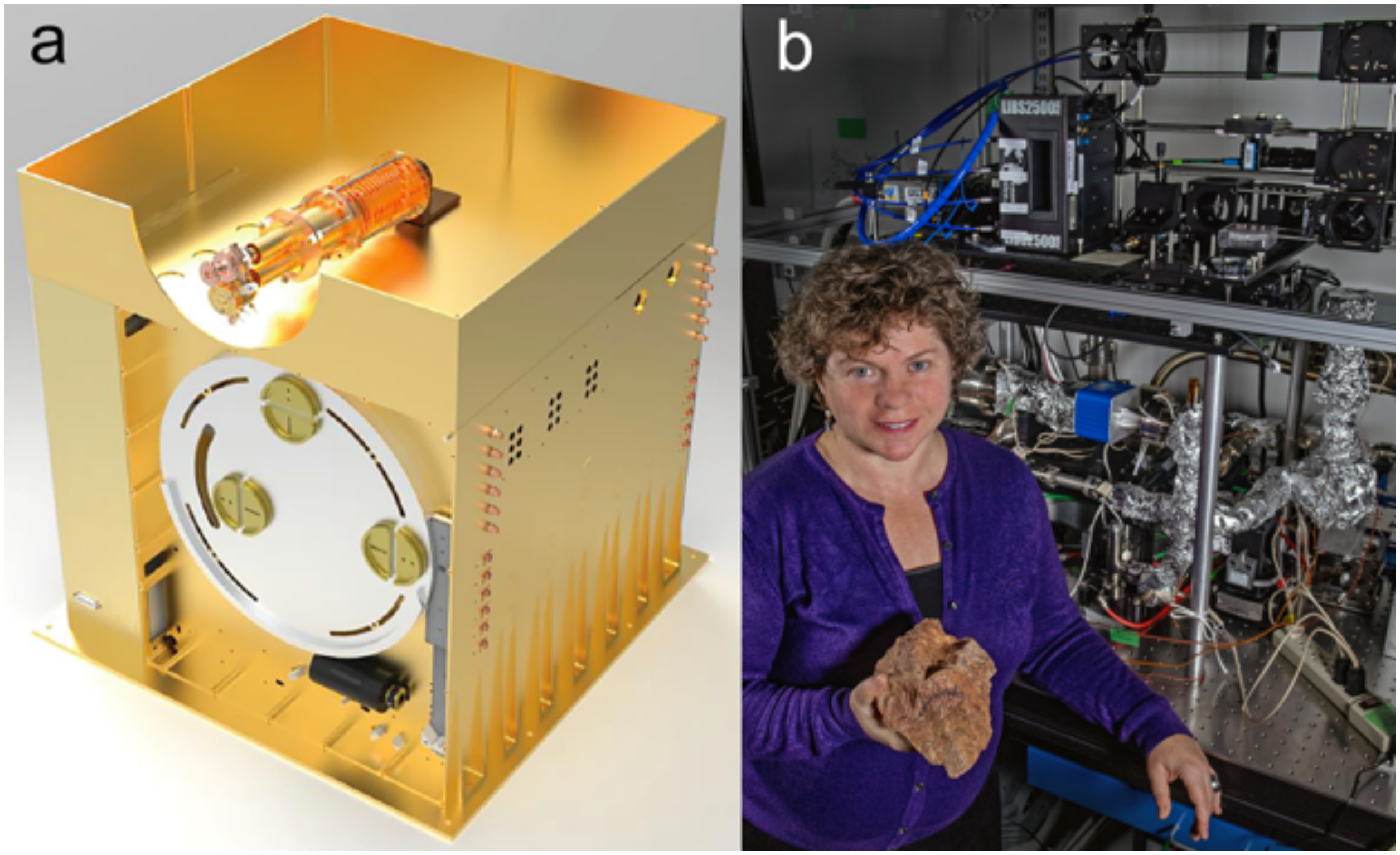

Figure 2: a) Rendering of the CDEX instrument (Southwest Research Institute); and b) KArLE instrument laboratory breadboard (GSFC).

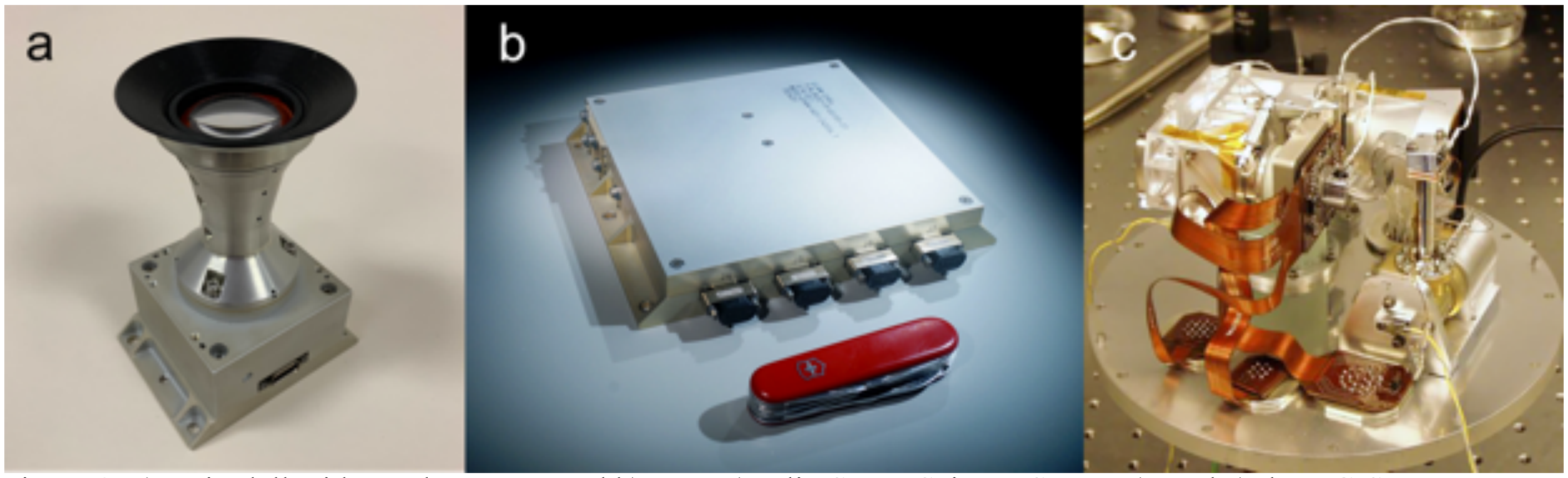

Figure 3: a) Heimdall wide-angle camera and b) DVR (Malin Space Science Systems); and c) the UCIS-Moon instrument (JPL).

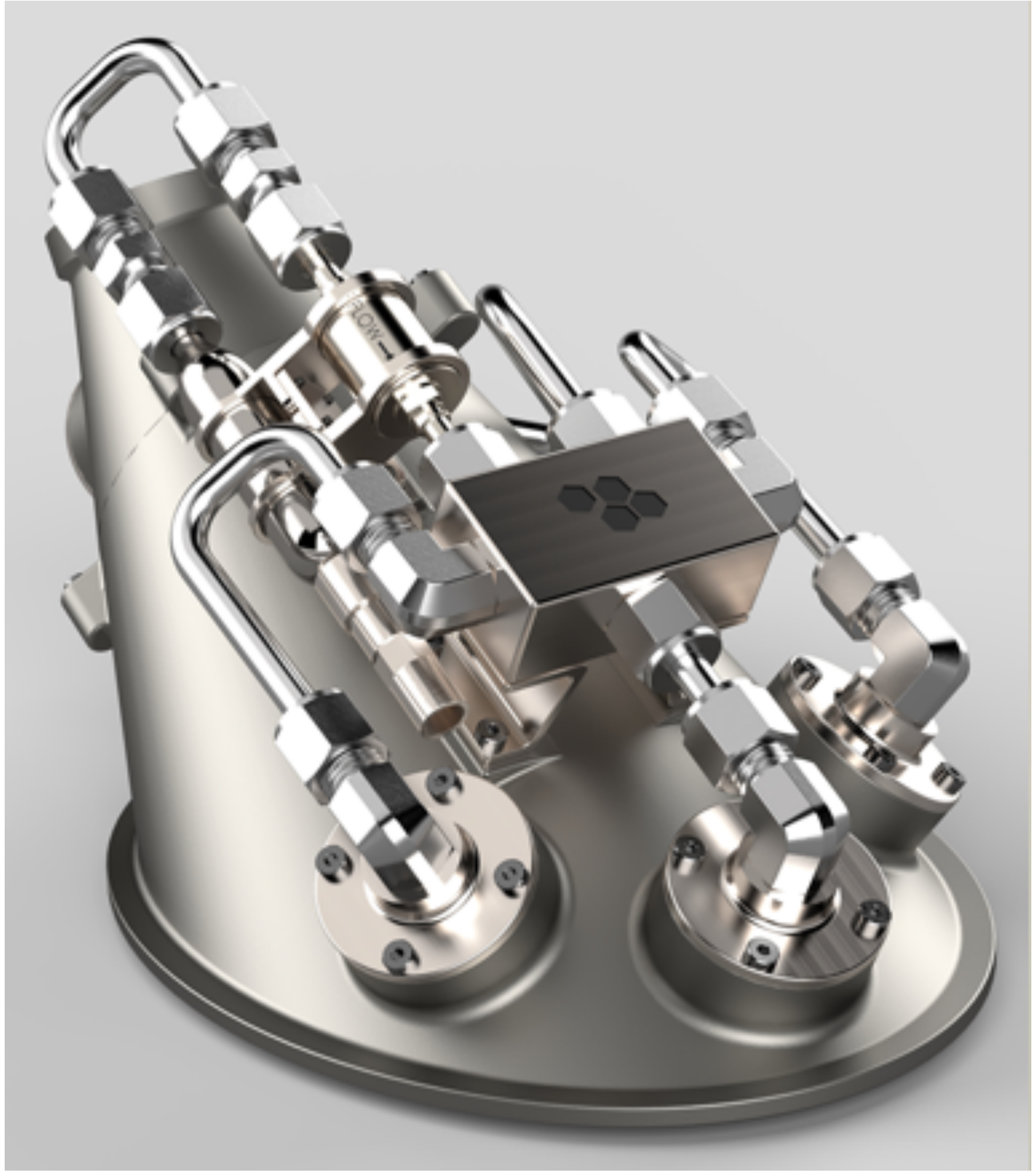

Figure 4: PlanetVac foot unit hardware (Honeybee Robotics).

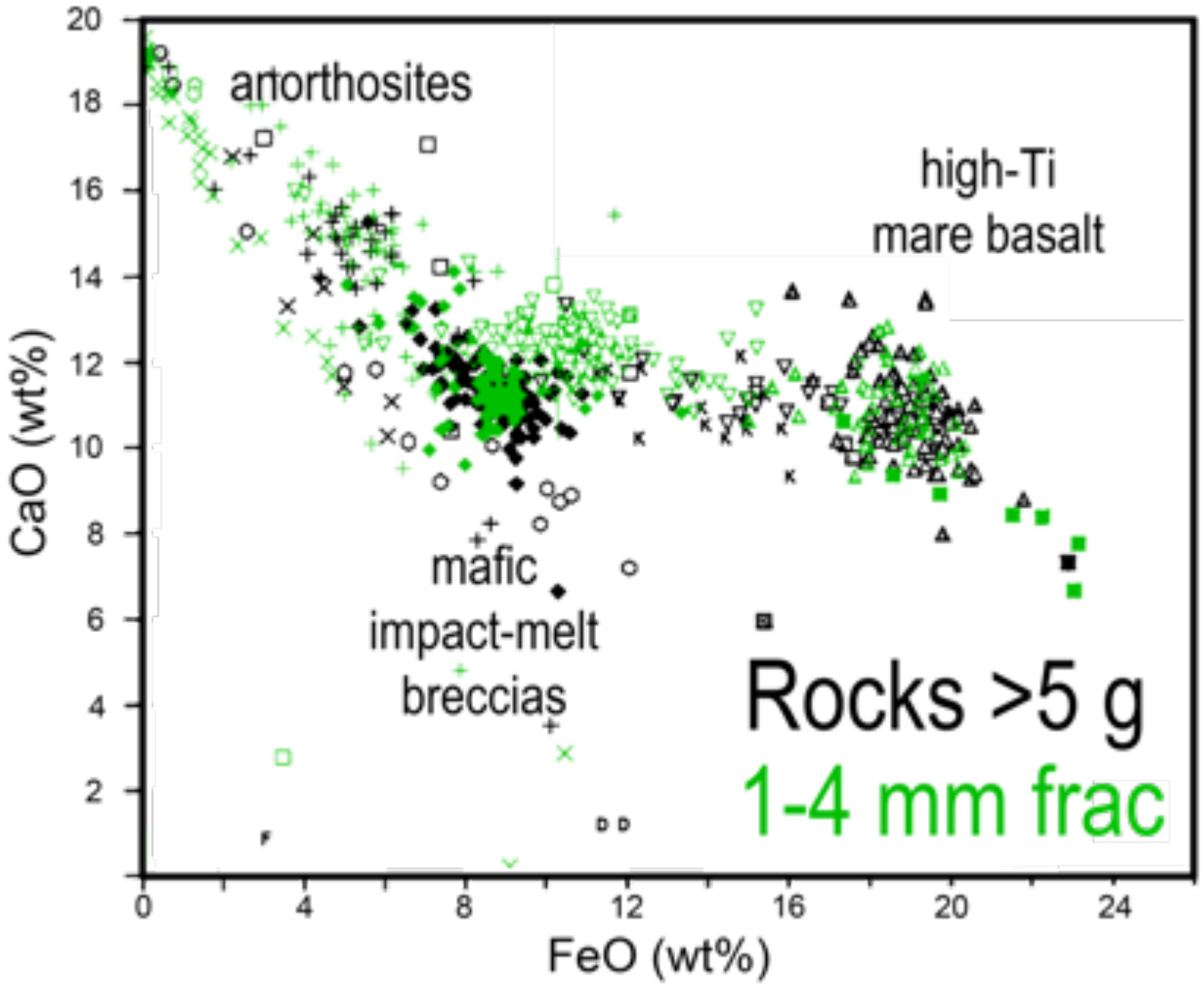

Figure 5: Samples derived from raking and sieving lunar regolith represent the complete geologic diversity of any individual site, as shown by the composition and range of 1-4 mm rocks sieved from rake and soil samples at the Apollo 17 site, which completely overlap that of individually-collected lunar rocks weighing >5g. Adapted from data in Jolliff et al. (1996).

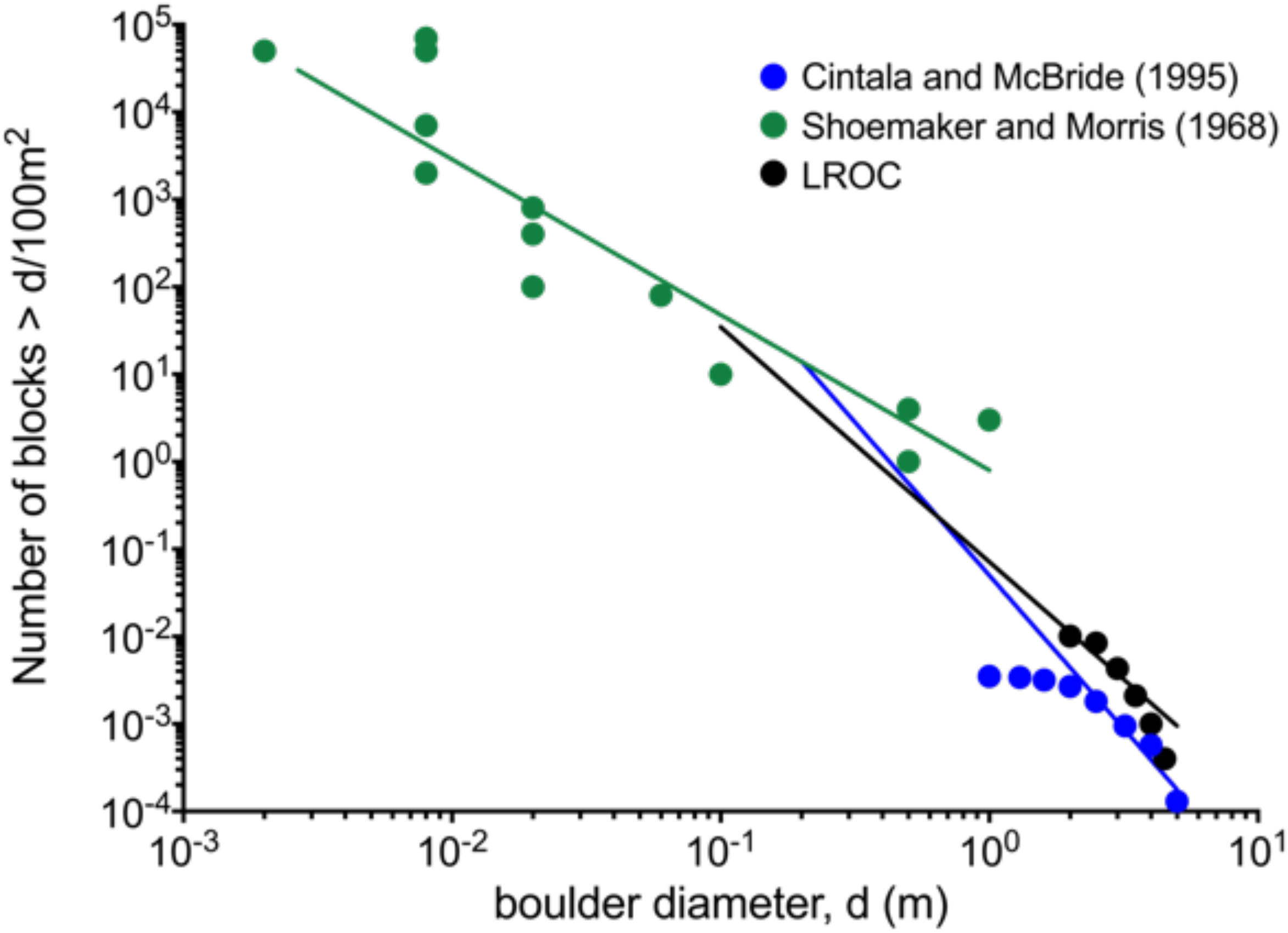

Figure 6: Comparison of grain size distribution estimates using data from the *Surveyor I* landing site. Roll-offs in block count using both *Surveyor* images and LROC NAC images are due to resolution limitations of the data.

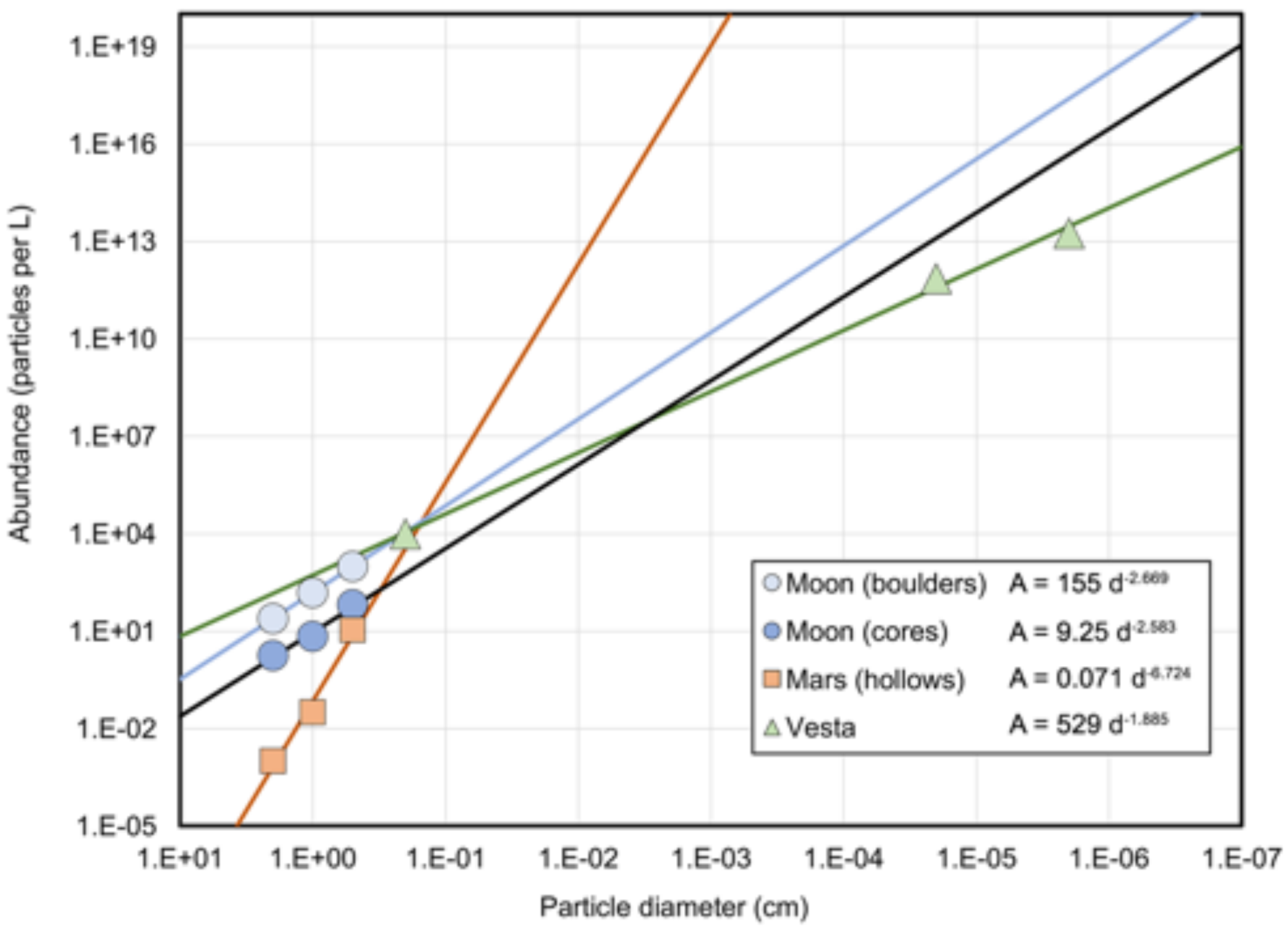

Figure 7: Compilation of grain size-frequency distributions for 0.5-2 cm rocks in the mature regolith of the Moon, Mars, and Vesta.

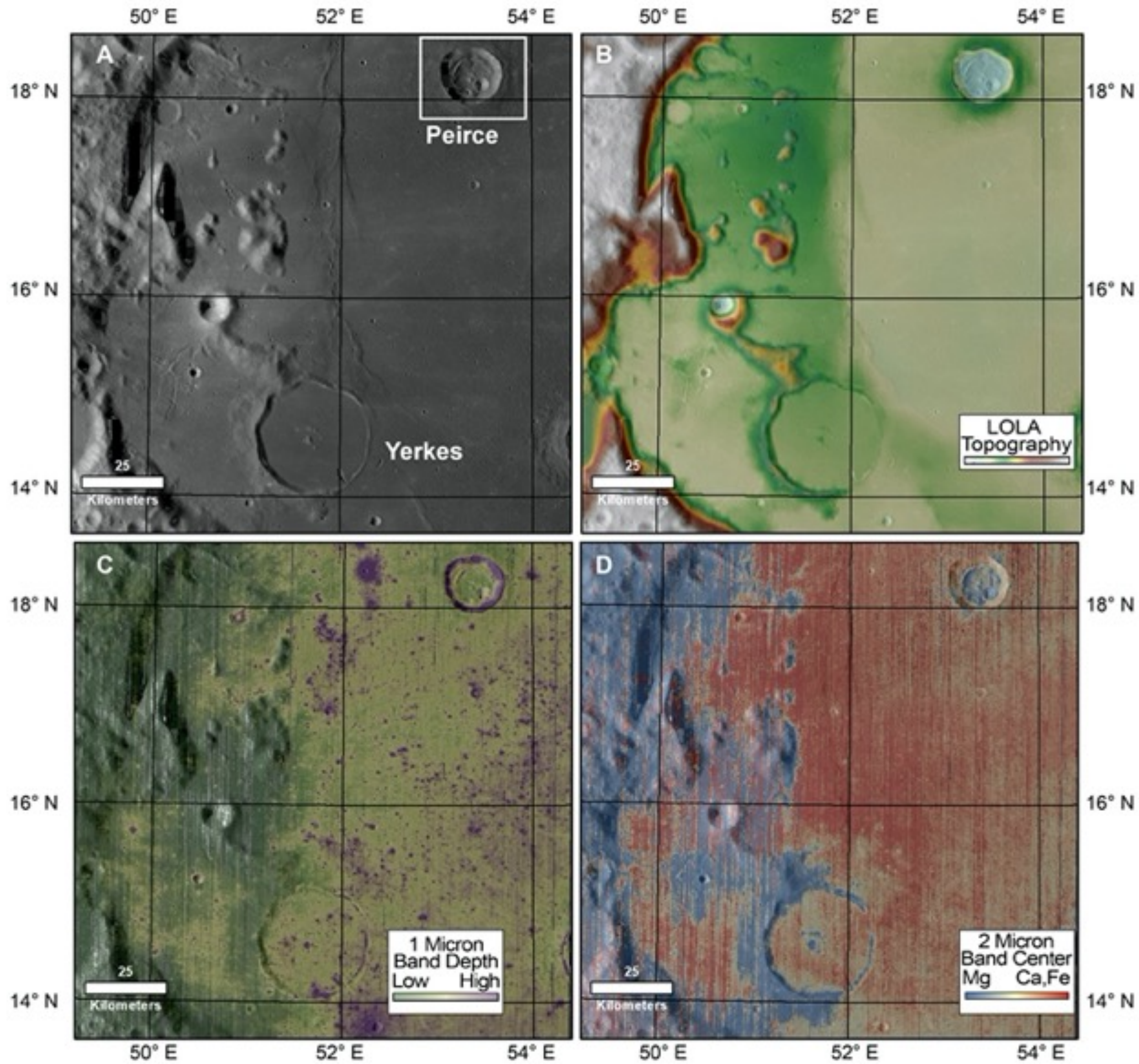

Figure 8: Peirce Crater geologic and mineralogic context in (A) *LROC* WAC imagery, (B) LOLA topography, (C) $M^3$ mafic abundance, and (D) *M*3 pyroxene composition parameters showing that Peirce excavated noritic Crisium impact melt, which is recognizably distinct from local mare basalts, enabling unambiguous discernment for in situ analyses.

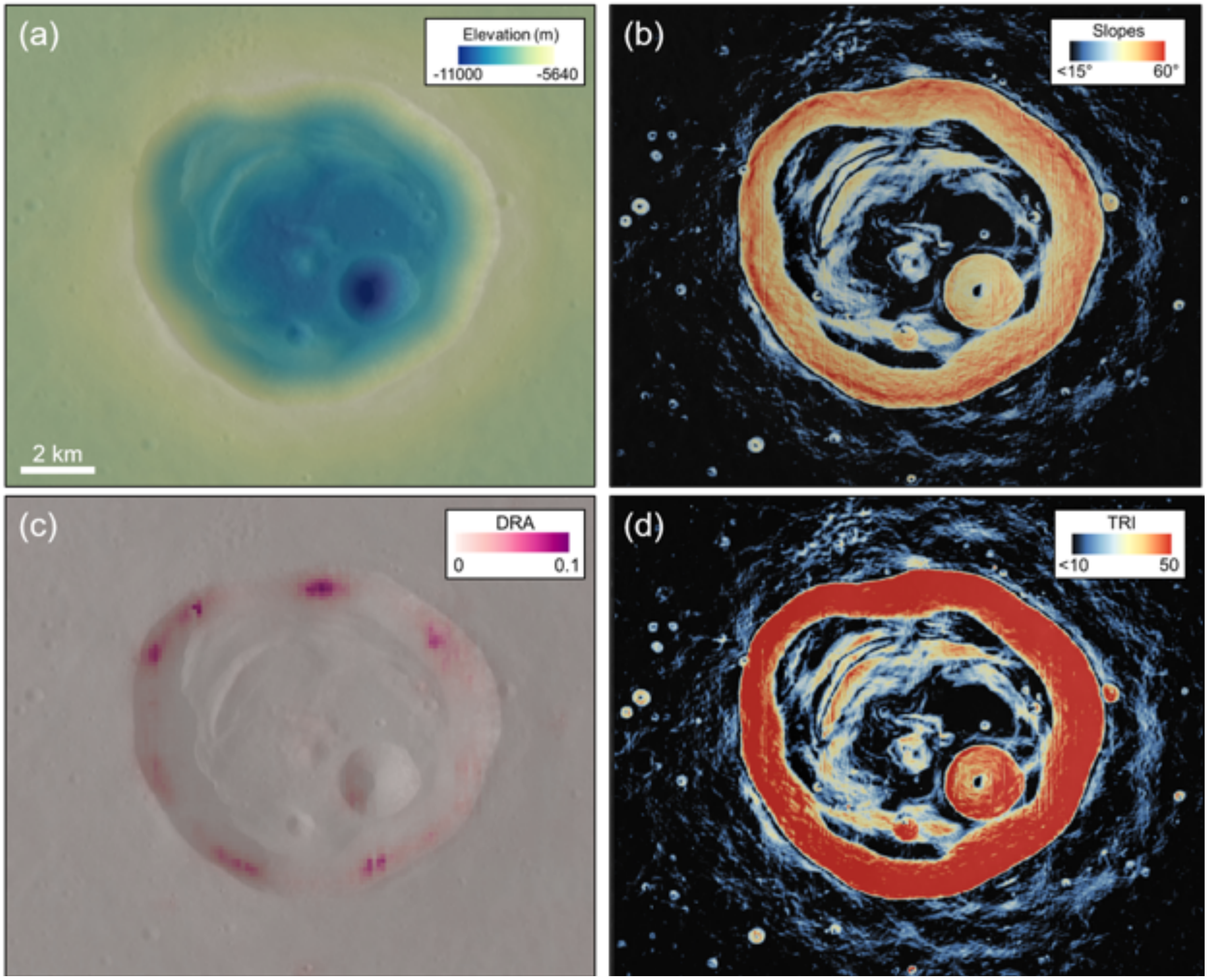

Figure 9: Landing-site safety evaluation for Peirce Crater: (a) LOLA elevation; (b) Slopes, with slopes < 15 deg (i.e., safe slope ranges) shown in black; (c) Diviner rock abundance (DRA); (d) Terrain Ruggedness Index (TRI), with safe values (TRI < 10) shown in black. A ~1K landing ellipse could be placed within the crater floor near a young, fresh crater.

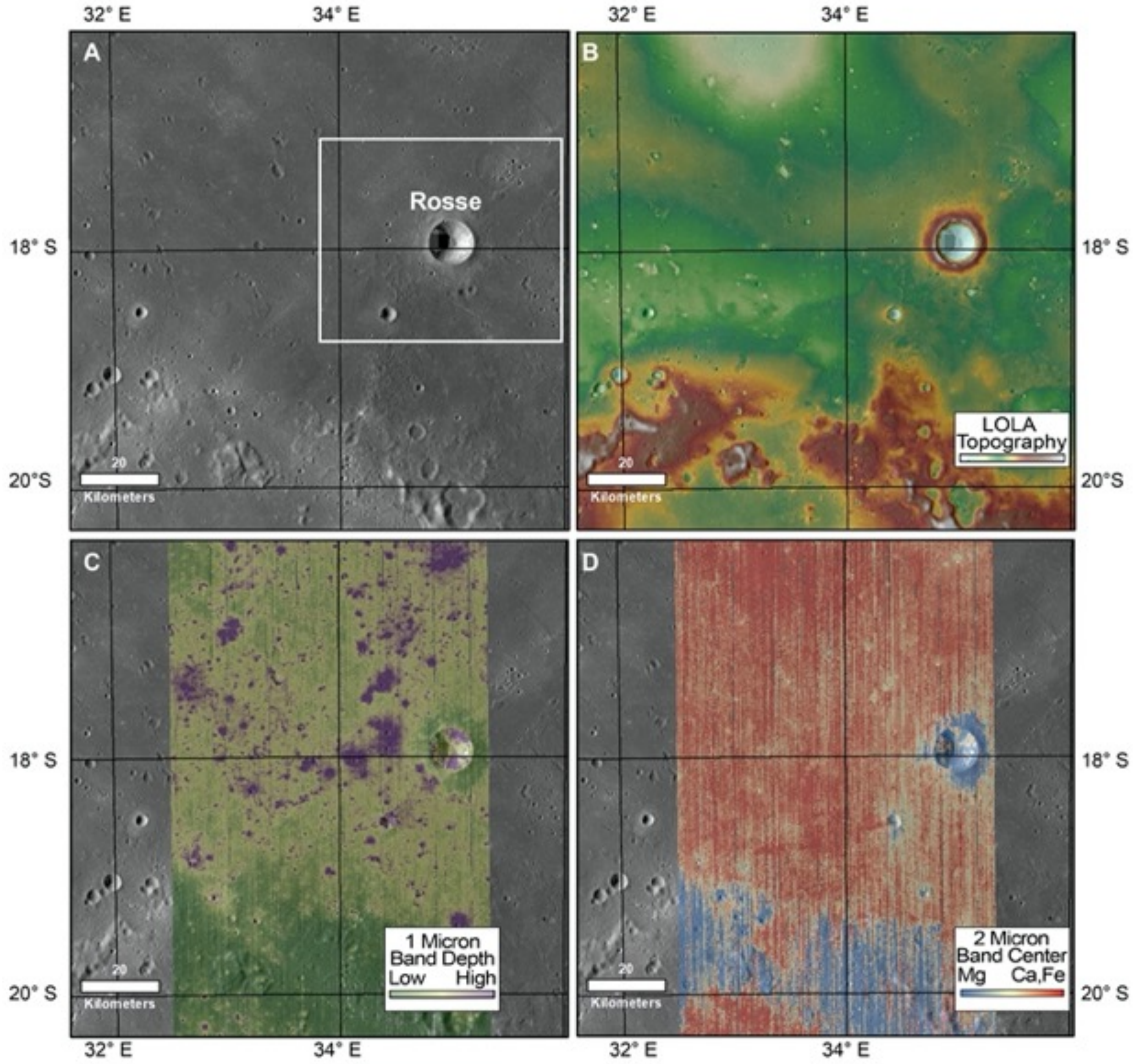

Figure 10: Rosse Crater geologic and mineralogic context in (A) LROC WAC imagery, (B) LOLA topography, (C) $M^3$ mafic abundance, and (D) $M^3$ pyroxene composition parameters showing that Rosse excavated noritic Nectaris impact melt from beneath the mare-flooded surface of Nectaris, enabling unambiguous discernment for *in situ* analyses.

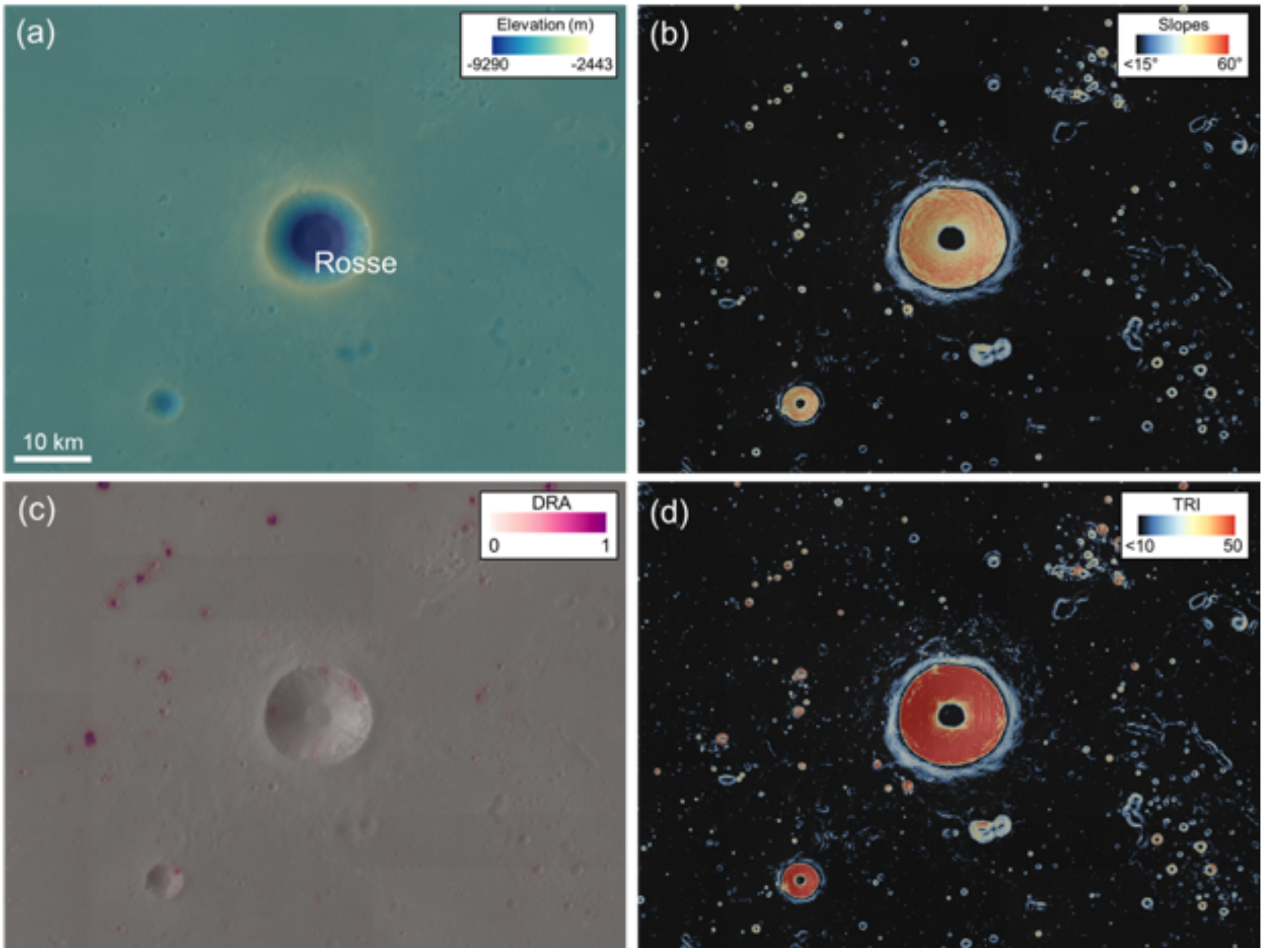

Figure 11: Landing-site safety evaluation for Rosse Crater: (a) LOLA elevation; (b) Slopes, with slopes < 15 deg (i.e., safe slope ranges) shown in black; (c) Diviner rock abundance (DRA); (d) Terrain Ruggedness Index (TRI), with safe values (TRI < 10) shown in black. Multiple potential landing sites exist near the crater rim where Rosse ejecta may have excavated the Crisium impact-melt sheet.

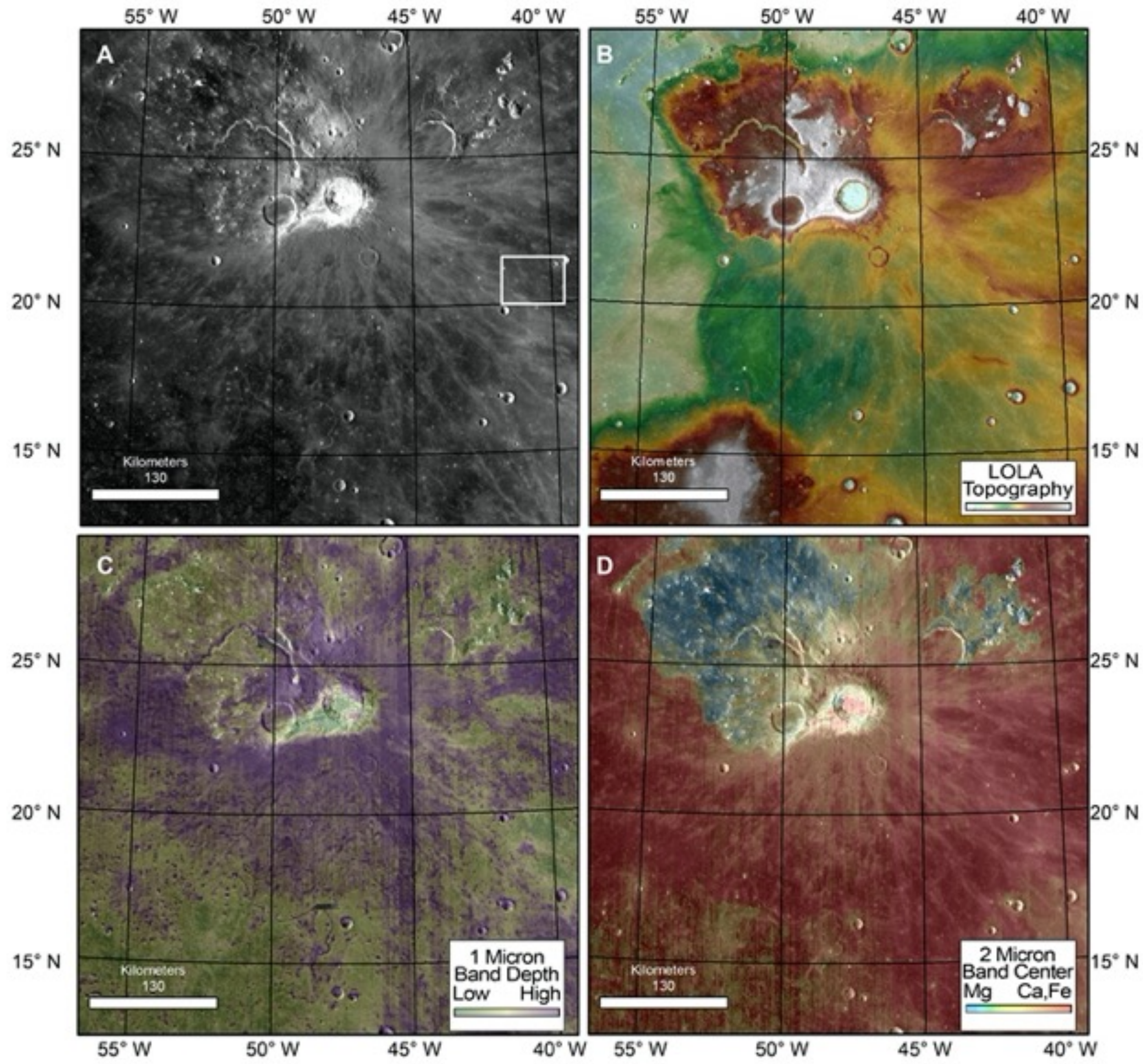

Figure 12: Geologic context for the P60 region: (A) LROC WAC imagery, (B) LOLA topography, (C) mafic mineral abundance, and (D) pyroxene composition showing that it exhibits a broad expanse of distinctly basaltic mineralogy, based on strong, relatively long-wavelength spectral absorption bands in $M^3$ data.

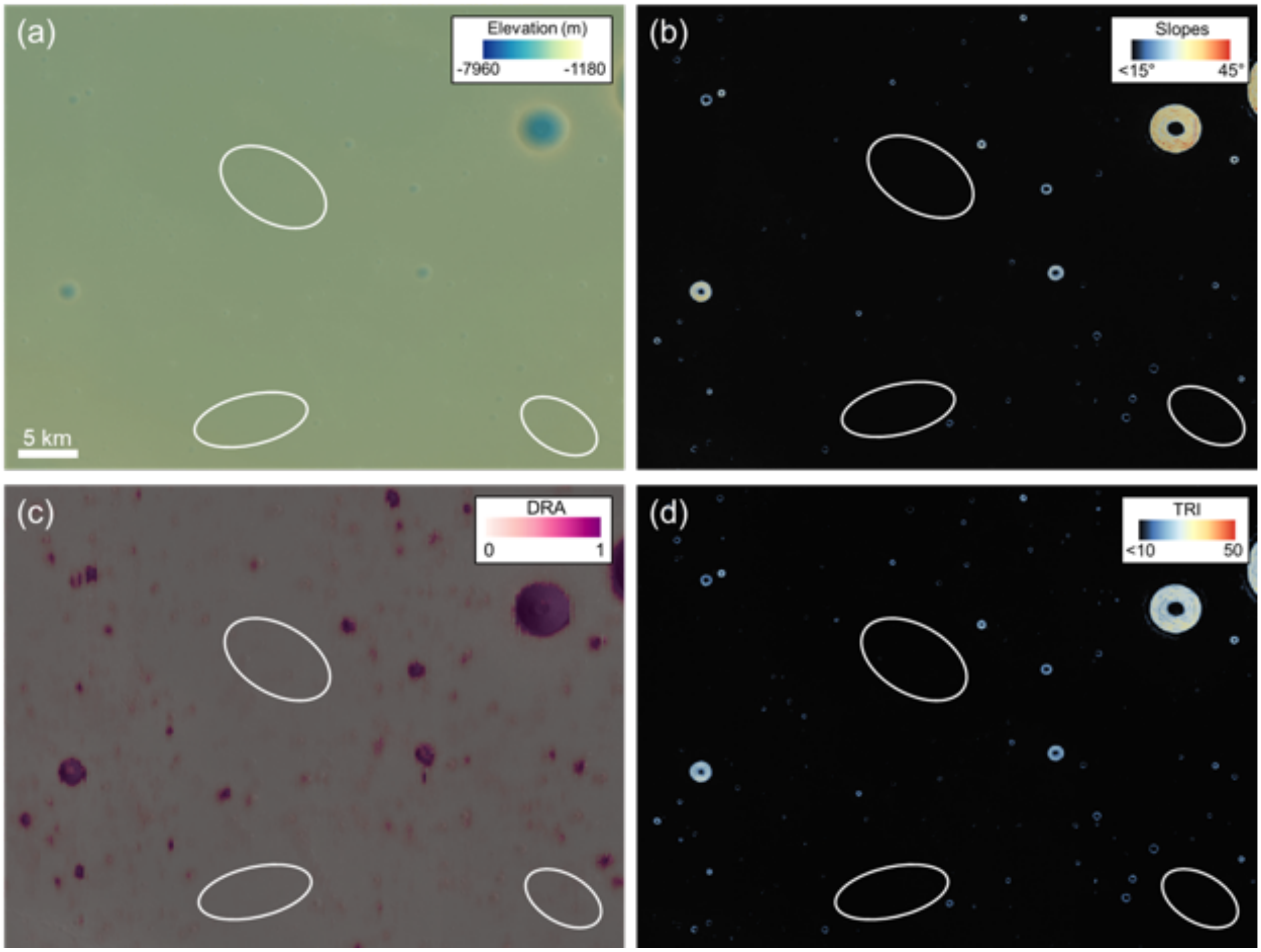

Figure 13: Potential landing areas in the P60 region, indicated by ellipses. Ellipse 1 is centered at 21.06°N, -40.43°E), Ellipse 2 is centered at (20.36°N, -40.70°E). (a) LOLA elevation; (b) Slopes, with slopes < 15 deg (i.e., safe slope ranges) shown in black; (c) Diviner rock abundance (DRA); (d) Terrain Ruggedness Index (TRI), with safe values (TRI < 10) shown in black.

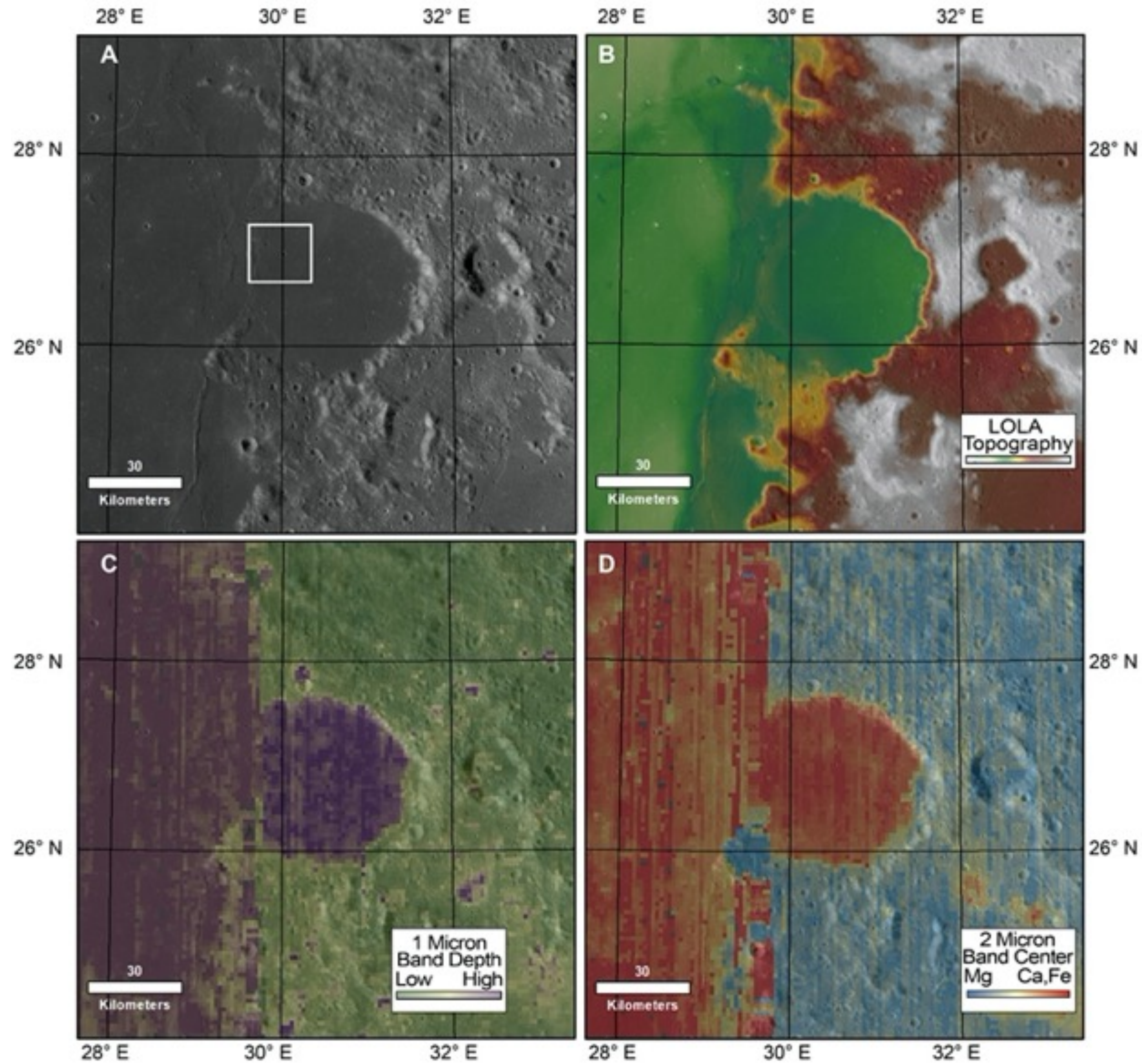

Figure 14: Geologic context for the Le Monnier crater region was assessed using (A) LROC WAC imagery and (B) LOLA topography. Mineralogical diversity, including (C) mafic mineral abundance and (D) pyroxene composition, was assessed using $M^3$ data. The mare floor of Le Monnier exhibits a distinctly basaltic mineralogy, based on strong, relatively long- wavelength spectral absorption bands in $M^3$ data.

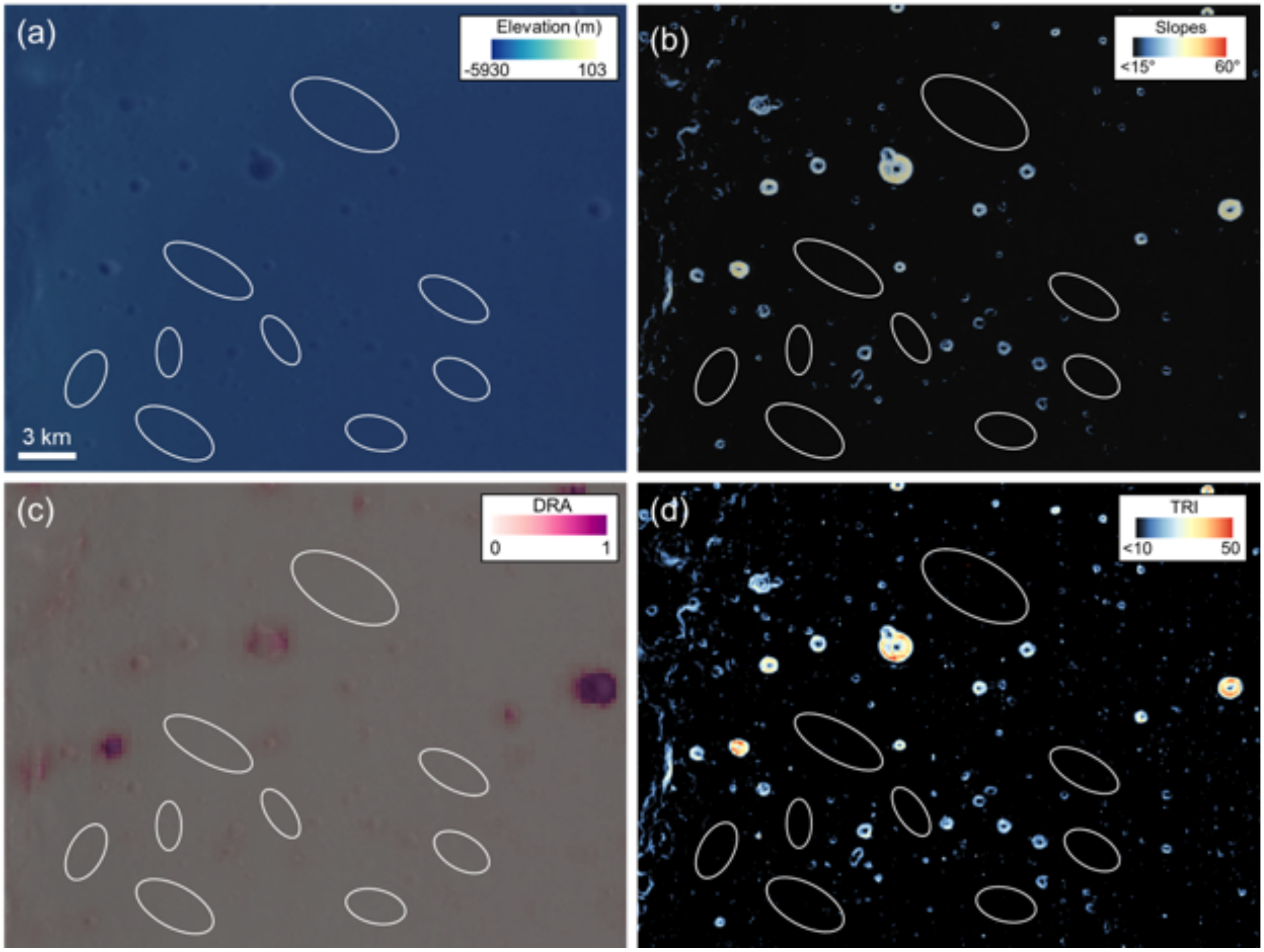

Figure 15: Potential landing areas inside Le Monnier crater, indicated by ellipses. (a) LOLA elevation; (b) Slopes, with slopes < 15 deg (i.e., safe slope ranges) shown in black; (c) Diviner rock abundance (DRA); (d) Terrain Ruggedness Index (TRI), with safe values (TRI < 10) shown in black. Image centered at (26.86°N, 30.08°E).

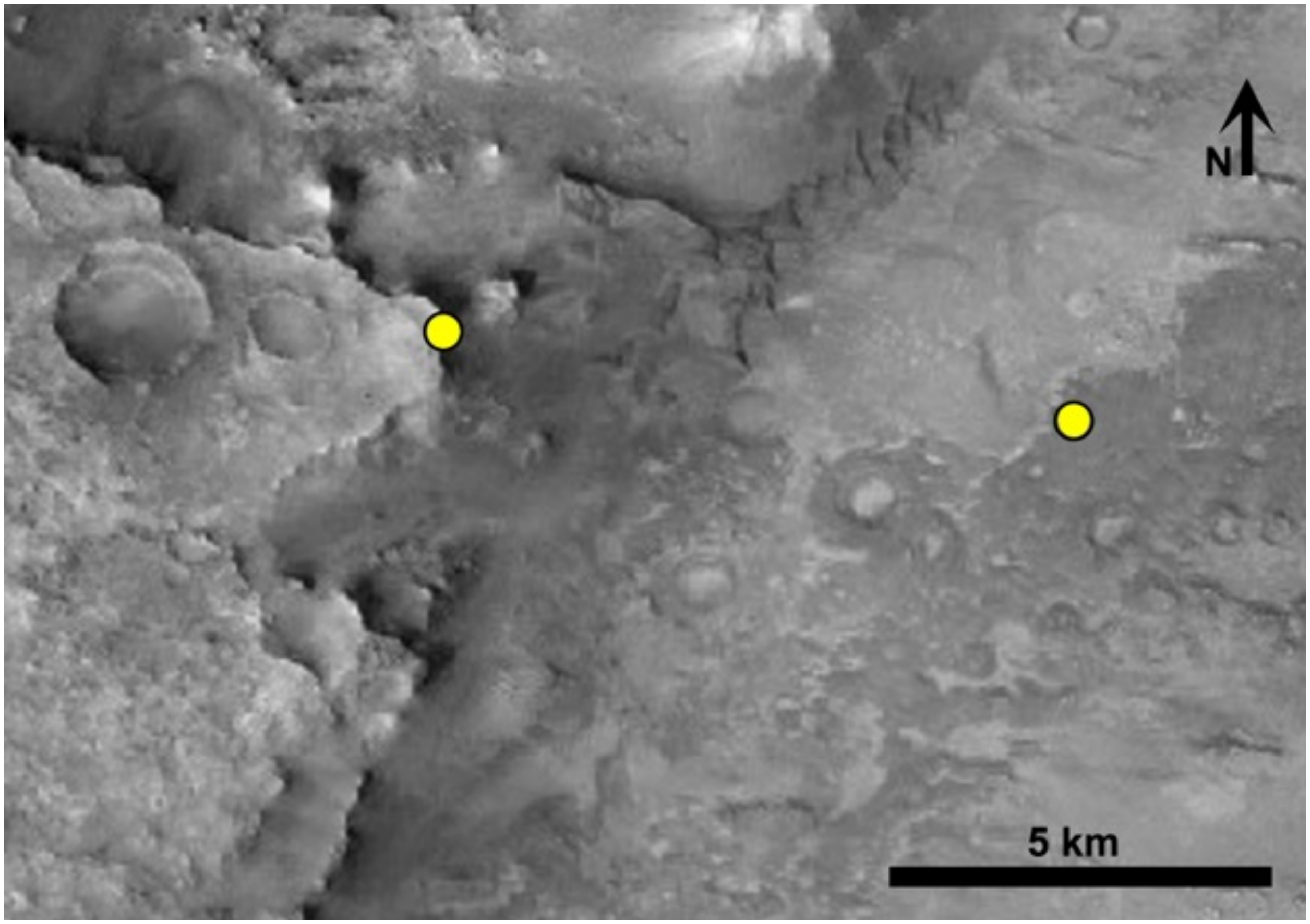

Figure 16: *Mars Reconnaissance Orbiter* Context Camera image of the floor of Nili Fossae Trough (right two thirds) and Noachian-aged crust (with orbital signatures of low-Ca pyroxene and clay) to the west (left one third), proposed as the landing site for both *Curiosity* and *Perseverance*. The yellow dots represent locations just under 5 km apart where Hesperian lava and clay minerals might be accessed.

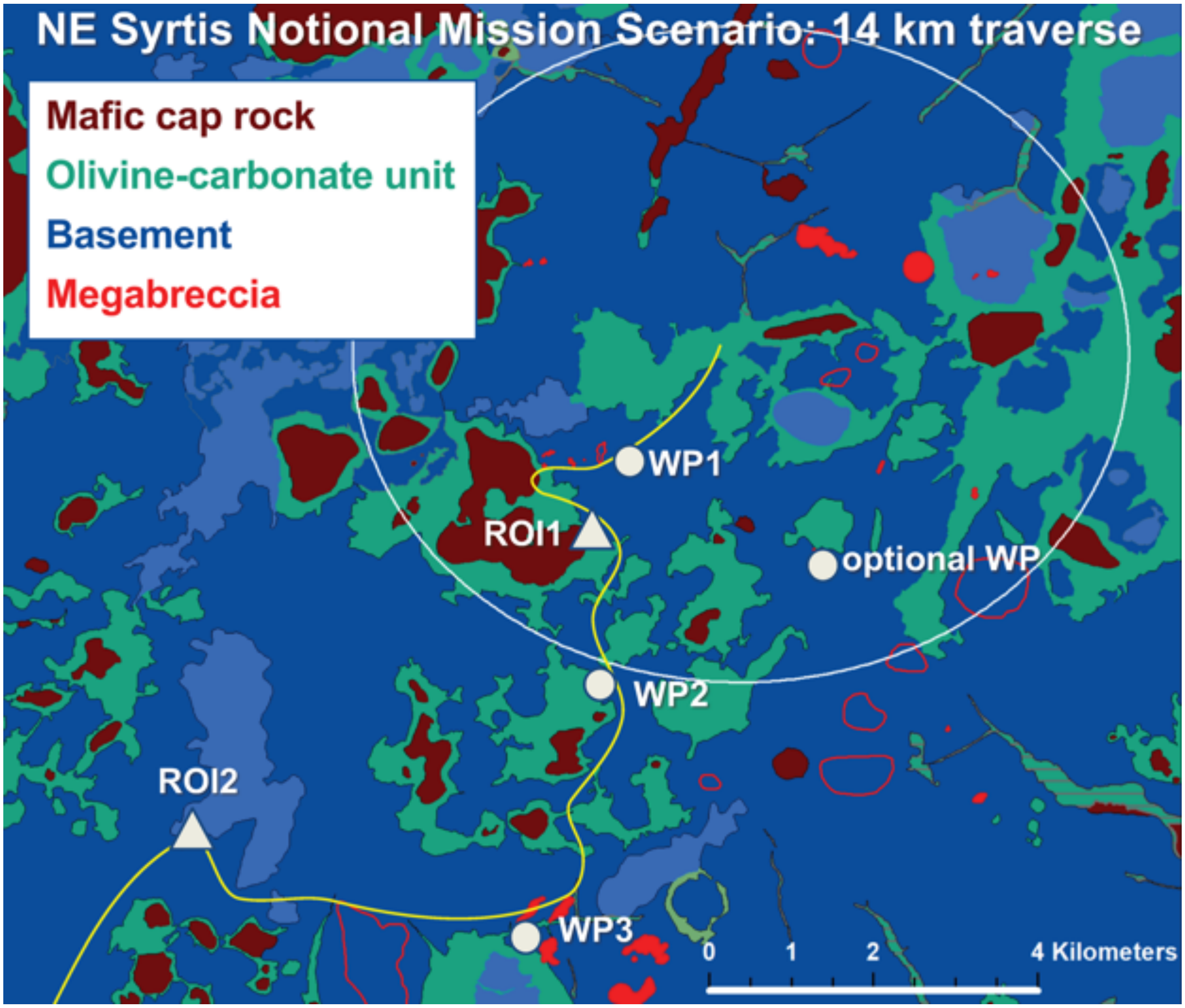

Figure 17: Northeast Syrtis Major lithologic map showing exposures of Noachian crust and mafic capping units. Adapted from Bramble et al. (2017) with candidate landing site ellipse for *Perseverance* (12×10 km) and notional traverse to regions of interest (ROI) as shown in Sun et al., (2018).

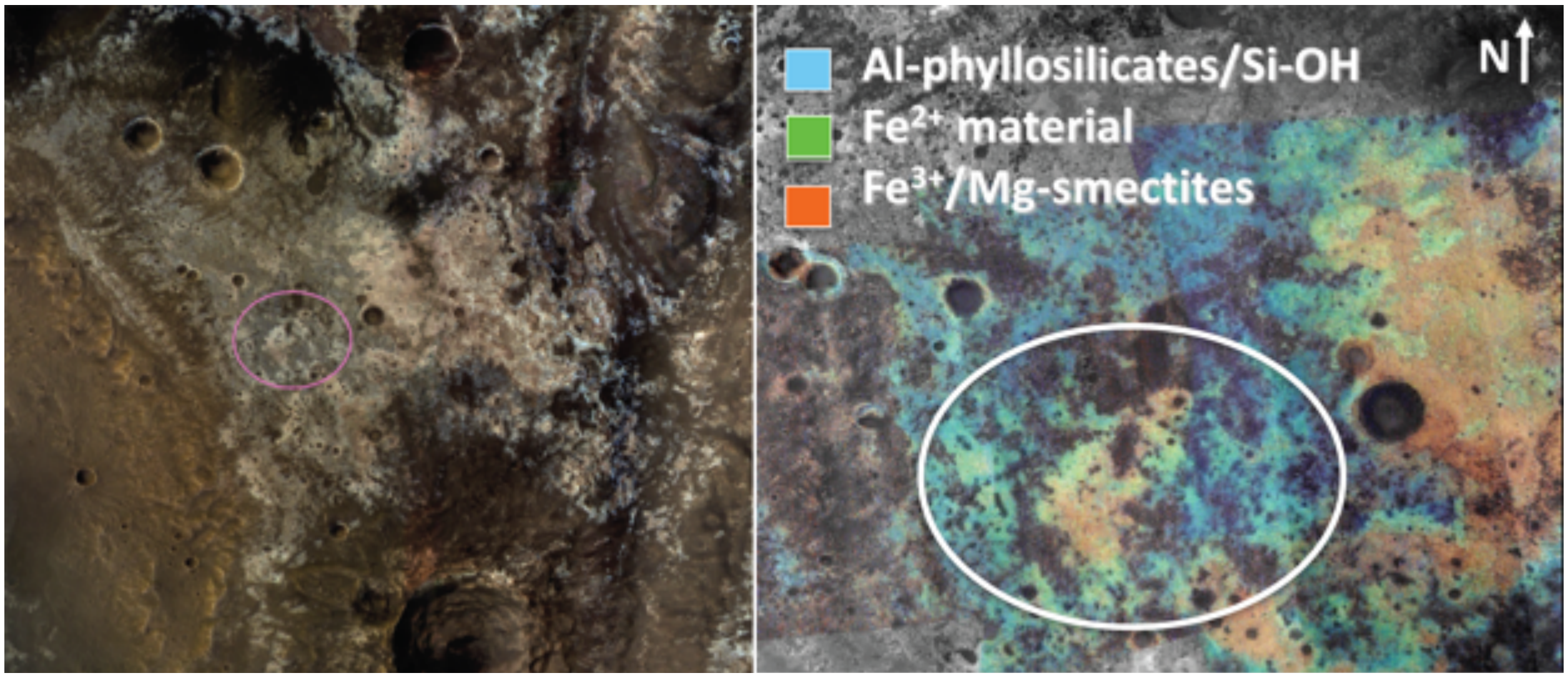

Figure 18: Mawrth Vallis landing site showing *Mars Express* High-Resolution Stereo Camera color (left) and mineralogy derived from *Mars Reconnaissance Orbiter* Compact Reconnaissance Imaging Spectrometer for Mars (right) (Bishop et al., 2008). The ellipse envelopes are candidate landing sites for *Perseverance* (12×10 km) as shown by Bishop et al., (2017).

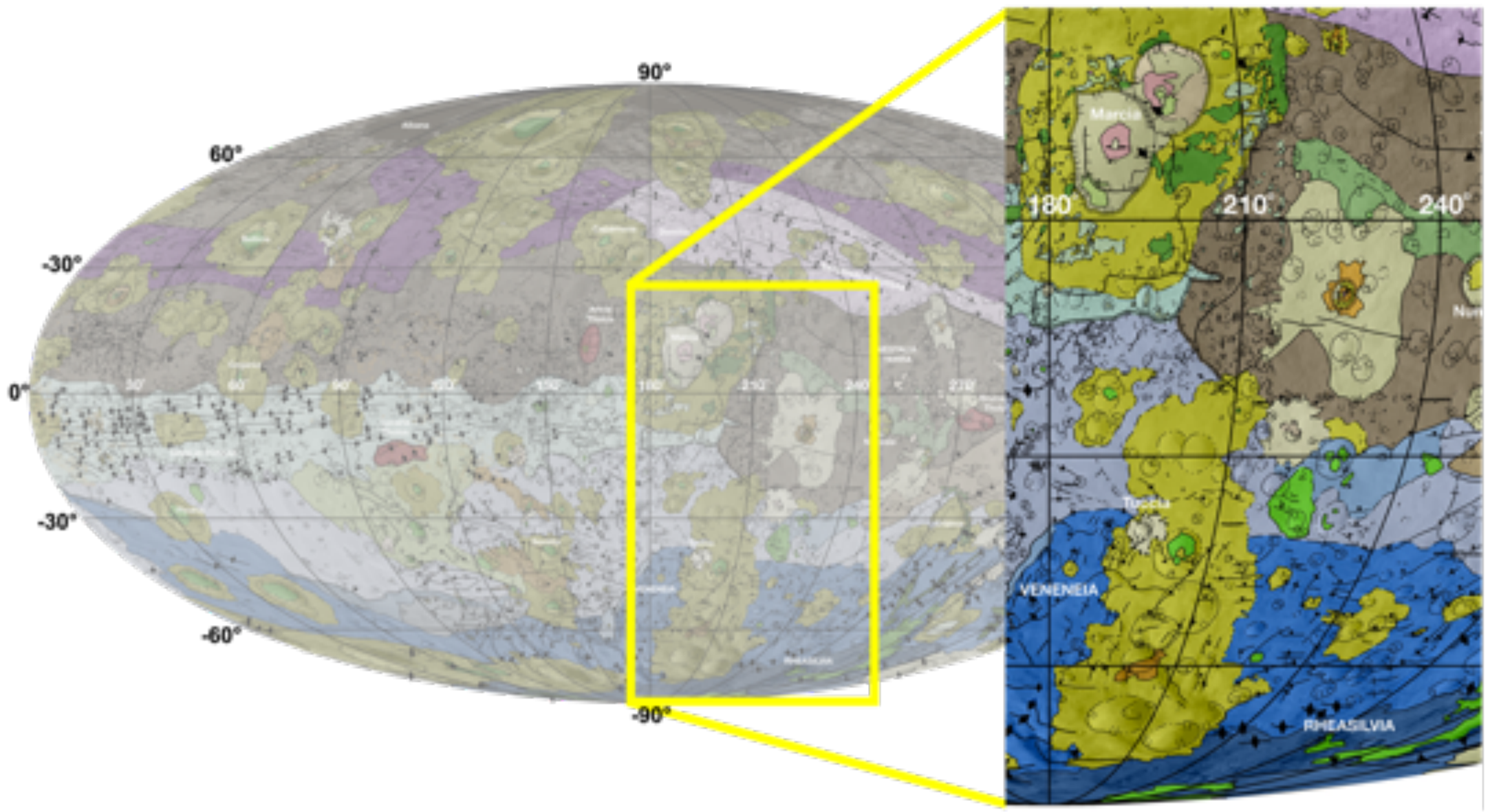

Figure 19: Excerpt from the geologic map of Vesta derived from Dawn spacecraft data, showing the Rheasilvia and Veneneia basins (dark blue colors) near the south pole, and Marcia crater (light brown and yellow) near the equator. Map is a Mollweide projection, centered on 180 degrees longitude using the Dawn Claudia coordinate system.

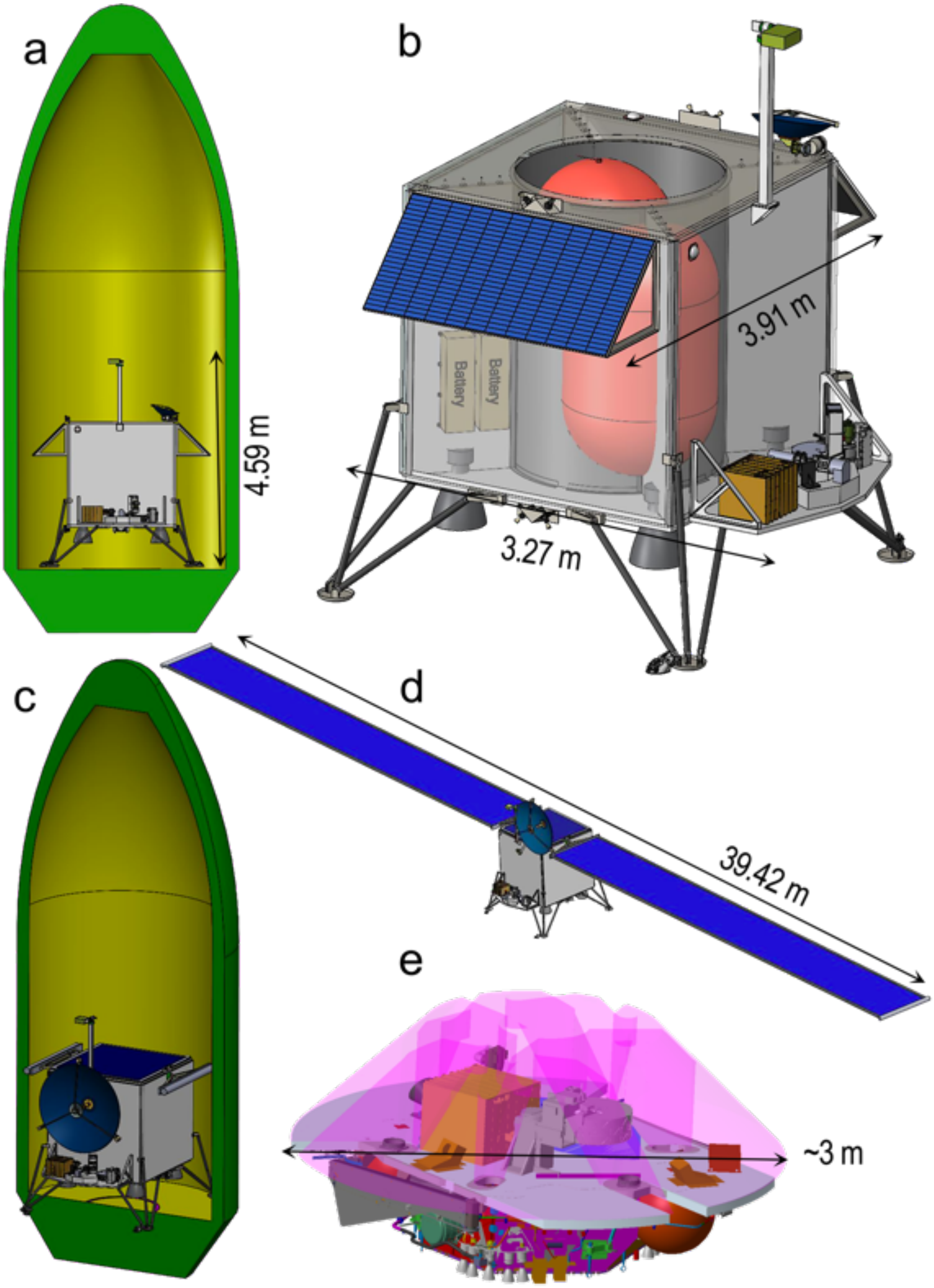

Figure 20: Views of the Moon, Vesta, and Mars geochronology landers. A) Lunar lander in Falcon 9 Heavy 5-m fairing; b) lunar lander structure accommodating the instrument deck; c) Vesta hopper with stowed solar arrays in a Falcon 9 Heavy 5-m fairing (lander dimensions are similar to the lunar lander) and d) Vesta lander with unfurled solar arrays; and e) Instrument deck for a *Phoenix*-like lander shown in a 3-m heat shield (pink envelope).

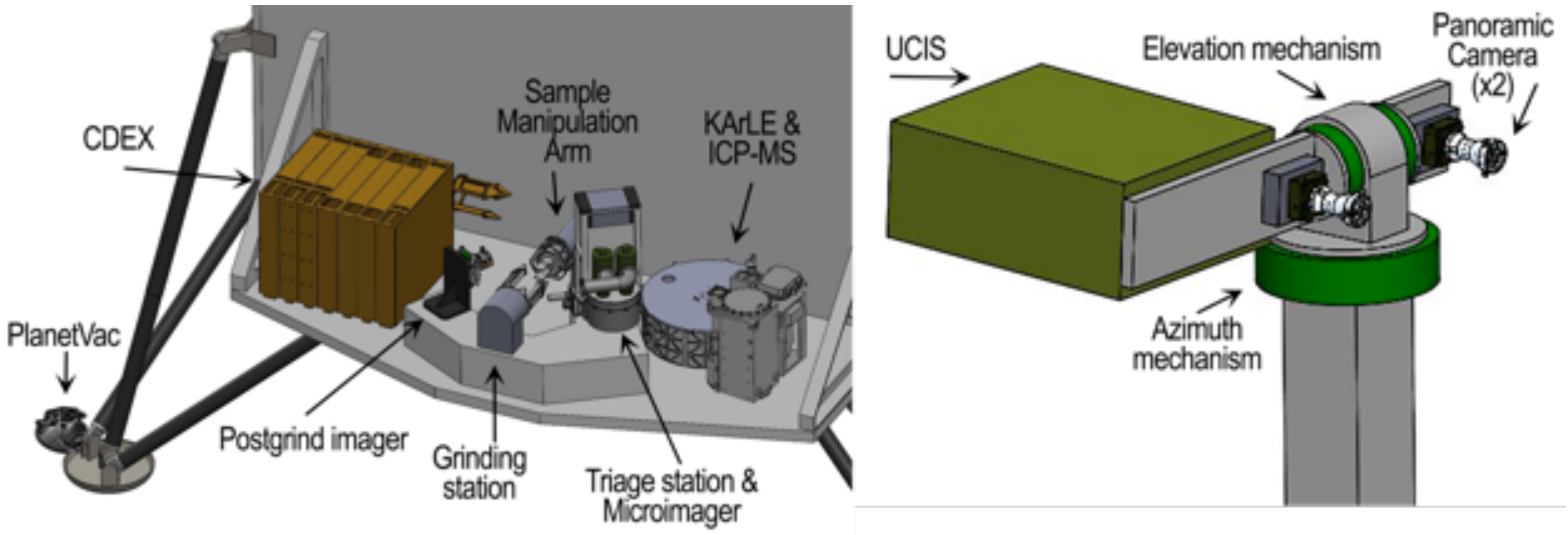

Figure 21: Notional geochronology payload layout for the Moon and Vesta missions.

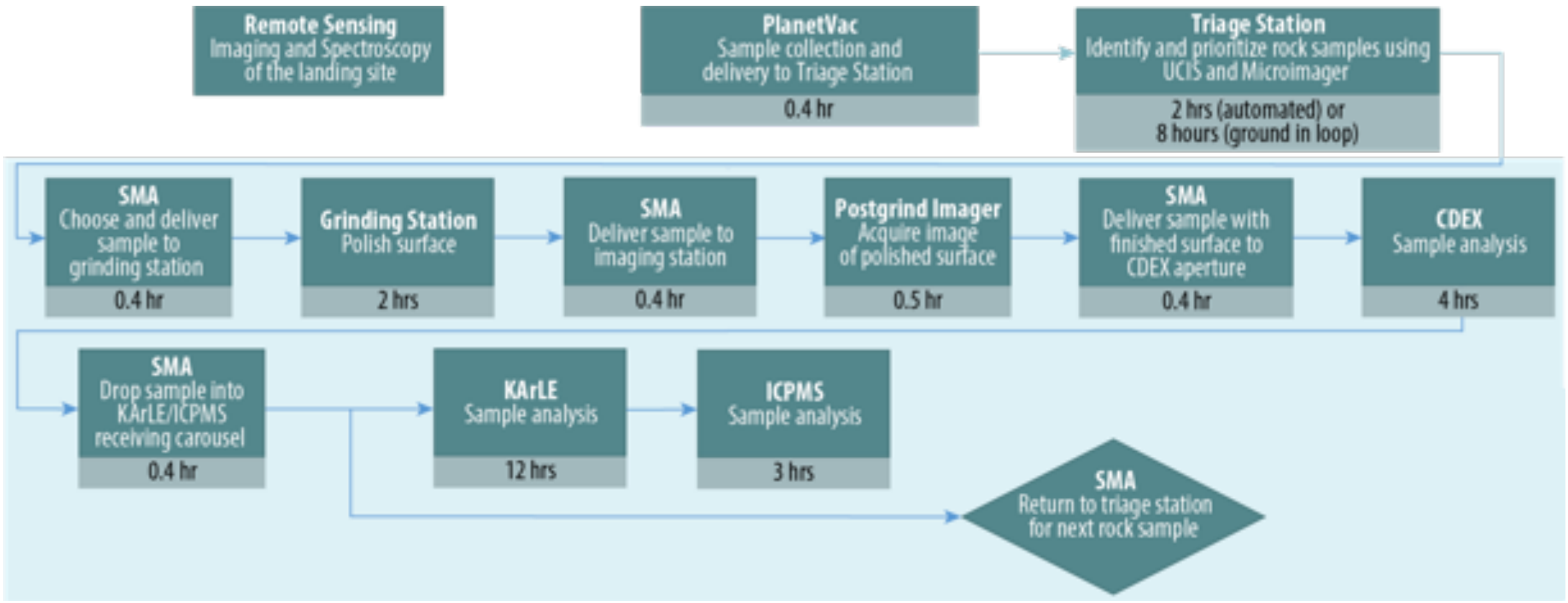

Figure 22: Notional sample analysis sequence common to all mission concepts. All times given are best estimates plus 100% margin.